\documentclass[twocolumn,trackchanges]{aastex631}
\usepackage{amsmath,amssymb}

\usepackage{color}
\usepackage{ulem}

\DeclareRobustCommand{\erase}{\bgroup\markoverwith{\textcolor{red}{\rule[.5ex]{2pt}{0.4pt}}}\ULon}

\received{}
\revised{}
\accepted{}
\submitjournal{ApJ}

\begin{document}
\title{The Impact of Stellar Radiative Feedback on Formation of Young Massive Clusters via Fast HI Gas Collisions}

\author{Ryunosuke Maeda}
\affiliation{Astronomical Institute, Graduate School of Science, Tohoku University, 6-3 Aramaki, Aoba, Sendai 980-8578, Japan}
\email{e-mail: maeda.ryunosuke@astr.tohoku.ac.jp}
\author{Tsuyoshi Inoue}
\affiliation{Department of Physics, Konan University, 8-9-1 Okamoto, Higashi-Nada, Kobe 658-8501, Japan}
\author{Kazuyuki Omukai}
\affiliation{Astronomical Institute, Graduate School of Science, Tohoku University, 6-3 Aramaki, Aoba, Sendai 980-8578, Japan}
\author{Yasuo Fukui}
\affiliation{Department of Physics, Graduate School of Science, Nagoya University, Furo-cho, Chikusa, Nagoya 464-8602, Japan}
\author{Kisetsu Tsuge}
\affiliation{Department of Physics, Graduate School of Science, The University of Tokyo, 7-3-1 Hongo, Bunkyo, Tokyo 113-0033, Japan}

\begin{abstract}
Young massive clusters (YMCs) are dense aggregates of young stars and are often speculated as potential precursors to globular clusters. 
However, the formation mechanism of massive and compact gas clumps that precede YMCs remains unknown.
In this paper, we study the formation of such massive clumps via fast HI gas collisions ($\sim 100~\mathrm{km~s^{-1}}$) as suggested by recent observations
and their subsequent evolution into YMCs by using three-dimensional magnetohydrodynamics simulations involving self-gravity and detailed thermal/chemical processes. 
In particular, the impact of ionization feedback from stellar radiation is included in an approximate fashion where the temperature within the HII regions is elevated to 10,000 K, while supernova feedback is not included. We examine whether the resulting massive clumps can survive this ionization feedback and evolve into YMCs.
Our simulations reveal the emergence of gas clumps that do not only possess substantial mass ($\sim 10^5~ \mathrm{M_\odot}$) but also sufficient compactness ($\sim 5~\mathrm{pc}$). 
Notably, these clumps exhibit significantly higher escape velocities compared to the sound speed of the HII region, indicating effective gravitational retention of gas against feedback-induced evaporation. Consequently, these conditions foster efficient star formation within the massive gas clumps, ultimately leading to their evolution into YMCs. 
We also perform simulations involving lower-velocity gas collisions, approximately $15~\mathrm{km~s^{-1}}$,  typical shock velocities induced by galactic superbubbles. 
In contrast to the high-velocity collisions, we find that molecular cloud formation does not occur in the case of $1~\mathrm{cm^{-3}}$ gas collision, while YMC formation is observed in the presence of denser gas of $10~\mathrm{cm^{-3}}$. 
However, the formation of YMCs requires compression periods exceeding $10~\mathrm{Myr}$ in these cases, indicating a potential preference for gas collisions driven by intergalactic interactions rather than galactic superbubbles for YMC formation. 
\end{abstract}

\keywords{} 

\section{Introduction}\label{sec:intro}
Young massive clusters (YMCs) are 
exceptionally dense stellar systems, with densities exceeding $10^3\ \mathrm{M_{\odot}\ pc^{-3}}$ \citep{longmore2014formation}. 
While YMCs are as dense as globular clusters, they are as young as open clusters in the Milky Way, suggesting they might be precursors of globular clusters \citep{portegies2010young}. 
Around a hundred YMCs have been observed in galaxies within the Local Group, while outside of it, there are galaxies that harbor several thousand YMCs \citep{longmore2014formation}. 
In addition, recent observations by the James Webb Space Telescope (JWST) suggest that early galaxies at redshifts of approximately $z \sim 4$ -- $10$ form at least $10$ -- $30\%$ of their stars within YMCs \citep[e.g.,][]{vanzella2022early,vanzella2023jwst,vanzella2023extremely,2024arXiv240103224A}. 
Additionally, numerical simulations indicate that massive star clusters are likely to form from warmer gas due to the combination of intense FUV radiation and low metallicity in the early universe
\citep{2023MNRAS.522.2495G,2024arXiv240304824S}. 
This demonstrates that understanding YMC formation is crucial for elucidating the star formation history of galaxies.

A variety of numerical simulations have been performed to study star cluster formation through the gravitational collapse of spherical turbulent molecular clouds \citep[e.g.,][]{dale2012ionizing, kim2018modeling, kim2021star, fukushima2020star, fukushima2022far, fujii2022sirius,he2019simulating,he2020simulating,grudic2018feedback,guszejnov2022cluster}. 
The results of these simulations indicate that a high star formation efficiency (SFE) can be achieved when starting with high surface density, massive compact gas clumps as initial conditions. \citet{fukushima2022far} demonstrate that a massive and dense star cluster resembling a YMC can form if the initial molecular clouds are as massive as $10^5\ \mathrm{M_\odot}$ and as compact as $200\ \mathrm{M}_{\odot}\ \mathrm{pc}^{-2}$. From these simulations, it is inferred that the formation of YMCs is feasible as long as a significant amount of gas can be accumulated in a sufficiently compact region. 
For the formation of these gas clumps, high external pressure \citep{elmegreen1997universal} and/or large velocity dispersion \citep{Fujii2015,fujii2016formation} appear to be necessary, as observed in regions such as cloud collisions, galactic mergers, and galactic centers \citep[see also,][]{longmore2014formation}. 
However, the exact mechanism behind the formation of YMC precursor clouds still remains a mystery.

Recent observational studies have focused on the gas structures surrounding YMCs. In the Large Magellanic Cloud (LMC), 
\cite{fukui2017formation} and \cite{tsuge2019formation} performed detailed analyses of neutral atomic hydrogen (HI) gas using data from the Australia Telescope Compacta Array and Parkes HI gas surveys \citep{staveley1997hi,kim1998hi,kim2003neutral}. They found characteristic structures indicative of HI gas collisions in the gas-surrounding regions where YMCs form: complementary spatial gas distributions and bridge features in velocity structures.
The colliding velocity is estimated to be $\sim 100\ \mathrm{km\ s^{-1}}$
\citep{fukui2017formation}, consistent with the velocity of gas collisions resulting from the galactic interaction with the Small Magellanic Cloud (SMC) \citep[]{fujimoto1990asymmetric, Bekki2007}. Observationally measured gas metallicity also suggests that the fast HI gas inflow observed in the LMC originates from the SMC \citep{fukui2017formation,tsuge2019formation,tsuge2024high}.
The scale of the gas colliding region is as large as $\sim \mathrm{kpc}$, and recent studies have obtained a three-dimensional (3D) structure of the colliding gas using dust extinction \citep{furuta2021three, furuta2022three}.
These findings suggest that fast HI gas collisions could potentially trigger the formation of YMCs in the LMC. Furthermore, YMC formation through fast gas collisions has also been reported in the Antennae Galaxies \citep{tsuge2021formation,tsuge2021formation2} and the M33 galaxy which is associated with the M31 galaxy \citep{tachihara2018}. 
These observations imply that YMC formation via gas collisions can be important in interacting galaxies in general.

Many HI gas collision simulations have been performed within the context of HI/molecular cloud formation \citep[e.g.,][]{inoue2008two,inoue2009two,inoue2012formation, Inoue2016, Koyama2000, Koyama2001,ballesteros1999turbulent,hartmann2001rapid,hennebelle2008warm,banerjee2009clump,heitsch2005formation,heitsch2008cooling,heitsch2006birth,heitsch2009effects,vazquez2007molecular,vazquez2006molecular,kobayashi2020bimodal,kobayashi2022nature,kobayashi2023metallicity,iwasaki2019early,iwasaki2022universal}. 
Those authors consider HI gas collision due to flows induced by supernovae within a galaxy, whose typical velocities and scales are respectively $\sim 10\ \mathrm{km\ s^{-1}}$ and $< 100\ \mathrm{pc}$.
Some studies suggest that the gravitational collapse of molecular clouds formed as a consequence of these collisions triggers the formation of star clusters with masses $M<10^4\ \mathrm{M_{\odot}}$ \citep[e.g.,][]{colin2013molecular,vazquez2017hierarchical,gonzalez2020effect}, comparable to the mass of open clusters but smaller than that of YMCs.

On the other hand, considering the faster ($\sim 100~\mathrm{km~s^{-1}}$) and larger-scale ($\gtrsim 100~\mathrm{pc}$) HI gas collisions observed around YMCs, \cite{maeda2021formation} demonstrated the formation of massive gas clumps with $M>2\times 10^5\ \mathrm{M_\odot}$ ($L\sim 6\ \mathrm{pc}$), which could serve as precursors to YMCs \citep[see also simulations of fast molecular cloud collisions, e.g.,][]{dobbs2020formation,dobbs2021properties}. These massive gas clumps resemble those used as initial conditions for YMC precursor clouds in \cite{fukushima2022far}, suggesting that fast HI gas collisions are a theoretically plausible mechanism for YMC formation. However, \cite{maeda2021formation} did not account for feedback effects from massive stars formed within the precursor massive clouds. Although the escape velocity of the formed gas clumps exceeded the sound speed of the HII region, indicating inefficient feedback, the formation of YMC precursor clouds should be verified by considering the feedback effect.

In this study, we conduct a series of simulations to investigate the formation of massive clusters via HI gas collision, while considering the feedback effects from massive stars. This paper is organized as follows: Section 2 introduces the simulation methods and setups. Section 3 presents the results of the simulations and provides an analysis. A discussion and summary of this study are provided in Section 4.
 
\section{Method} \label{sec:method} 
We locally simulate the gas collision region using a cubic numerical domain with a scale of $L_{\mathrm{box}} = 100\ \mathrm{pc}$ to ensure adequate resolution of the YMC scale ( $\lesssim 10\ \mathrm{pc}$). The numerical domain is discretized into $512^3$ uniform cells, resulting in a numerical resolution of approximately $\sim 0.2\ \mathrm{pc}$.
To understand the dynamics of the interstellar medium (ISM), we utilize 3D ideal magnetohydrodynamics (MHD) simulations using the Eulerian scheme, accounting for chemical reactions, radiative cooling/heating, self-gravity, and stellar feedback. 
As explained in Section 2.2, our simulations incorporate the formation of star cluster particles and include the ionization feedback from these particles by modeling it as heating within the HII regions. 
Note that our calculations do not take into account the feedback from supernova explosions; this impact is discussed in Section 4. 
Most of the fundamental equations and numerical techniques are outlined in our previous paper \citep{maeda2021formation} and the large part of the numerical code has been developed in \cite{inoue2008two,inoue2009two,inoue2012formation, Inoue2015, Inoue2016}.  
In this study, we added stellar feedback effects to the basic equations considered in \cite{maeda2021formation}. 
This section provides a brief overview of the basic equations and numerical methods detailed in our previous paper, along with a detailed explanation of the feedback effects incorporated in this study. 

\subsection{Basic equations}
The basic equations are as follows:  
\begin{equation}
\frac{\partial n_{\alpha}}{\partial t}+\nabla \cdot(n_{\alpha} \boldsymbol{v})=f_{\mathrm{\alpha}}\left(n_{\mathrm{\beta}}, N_{\mathrm{\beta}}, T, G_0 \right)-\dot{n}_\mathrm{\alpha, SF},
\end{equation}

\begin{align}
  \begin{split}
    \frac{\partial}{\partial t} \left(\rho \boldsymbol{v} \right)+\nabla \cdot  &\left( \rho \boldsymbol{v} \otimes \boldsymbol{v} + p -\frac{1}{4 \pi} \boldsymbol{B} \otimes \boldsymbol{B} \right. \\
    & \left. +\frac{1}{8 \pi} B^2 \right)=-\rho \nabla \Phi,
  \end{split}
\end{align}
\begin{align}
  \begin{split}
    \frac{\partial}{\partial t} (\frac{1}{2} \rho v^{2} &+\rho \epsilon+\frac{B^{2}}{8 \pi})+\nabla \cdot \left[\rho \boldsymbol{v}\left(\frac{1}{2} v^{2}+h\right) \right. \\
    & \left.+\frac{1}{4 \pi}(\boldsymbol{B} \times \boldsymbol{v}) \times \boldsymbol{B}\right]=-\rho \mathcal{L}\left(n_{\mathrm{\beta}}, N_{\mathrm{\beta}}, T, G_0\right)+S_\mathrm{MS},
  \end{split}
  \label{eq:engy-feed}
 \end{align}
\begin{equation}
\epsilon=\frac{1}{\gamma-1} \frac{p}{\rho}\ , \ h=\frac{\gamma}{\gamma-1} \frac{p}{\rho},
\end{equation}
\begin{equation}
\frac{\partial \boldsymbol{B}}{\partial t}+\mathbf{\nabla} \times(\boldsymbol{B} \times \boldsymbol{v})=0,
\end{equation}
\begin{equation}
\nabla \cdot \boldsymbol{B}=0,
\end{equation}
\begin{equation}
\nabla^{2} \Phi=4 \pi G \rho + 4 \pi G \rho_\mathrm{SC},
\end{equation}
where $\rho,\ p,\ T,\ \boldsymbol{v}$, $\boldsymbol{B}$, and $\Phi$ represent the density, pressure, temperature, velocity, magnetic field, and gravitational potential of the gas; $\epsilon$ is the specific internal energy, $h$ the specific enthalpy, $\gamma$ the specific heat ratio, and $G$ the gravitational constant. 

We consider several chemical species, and the following equation defines the density: 
\begin{equation}
\rho=\sum_{\mathrm{\alpha}} m_{\mathrm{\alpha}} n_{\mathrm{\alpha}}. 
\end{equation}
The subscripts $\alpha$ and $\beta$ in the above equations denote the chemical species $\mathrm{p},\ \mathrm{H},\ \mathrm{H}_{2},\ \mathrm{He},\ \mathrm{He}^{+},\ \mathrm{C},\ \mathrm{C}^{+},$ and $\mathrm{CO}$. 
In eq. (1)-(8), $n_{\alpha}$, $m_{\alpha}$, $N_\beta$, $f_{\mathrm{\alpha}}$, 
and $\mathcal{L}$ is the number density, the mass of molecules, the column density, the chemical reaction, and the net cooling rate per unit volume, respectively. 
The $G_0$ in the cooling rate is the background ultraviolet (UV) field strength. 
We include realistic chemical reactions and cooling/heating processes to follow the formation of molecular clouds from an optically thin HI medium \citep[see Tables 1 and 2 in][for more details]{inoue2012formation}. The effects of UV shielding in chemical reactions and heating/cooling processes are handled as described by \cite{inoue2012formation}, where the chemical reaction rates of $\mathrm{H_2}$ and $\mathrm{CO}$ molecules are calculated using an approximate radiative transfer method, specifically a two-ray approximation from the boundaries by using the optical depth of UV radiation. When calculating the chemical reaction rates, we consider UV radiation only from the boundaries and not from the formed massive stars.

We consider star formation and photoionization feedback as described in Section 2.2.  
The $\dot{n}_\mathrm{\alpha, SF}$ and $S_\mathrm{MS}$ in basic equations mean the star formation rate and the heating rate from massive stars. 
We follow the star cluster particle trajectories by solving the following equation: 
\begin{equation}
m_{\mathrm{SC}, \delta} \frac{d^2 \boldsymbol{r}_{\mathrm{SC}, \delta}}{d t^2} = -m_{\mathrm{SC}, \delta} \nabla  \Phi \label{eq:trac},
\end{equation}
where $m_\mathrm{SC}$, $\boldsymbol{r}_\mathrm{SC}$, and subscripts $\delta$ denote the star cluster particle mass, the position of the star cluster particle, and each of the formed star cluster particles, respectively. 
In this simulation, we incorporate the gravitational potential of the star cluster by redistributing its mass to the cells in which it resides ($\rho_\mathrm{SC}$).

We solve the above set of equations using the operator-splitting technique. 
The MHD part of the equations is solved in a conservative fashion using a second-order Godunov-type finite-volume scheme \citep{van1997flux} developed by \cite{sano1999higher}. 
We employ the consistent method of characteristics with constrained transport to solve the induction equation \citep{clarke1996consistent}. 
We treat the cooling/heating by way of a second-order explicit scheme and solve chemical reactions by using the piecewise exact solution method developed by \cite{inoue2008two}. 
The Poisson equation of self-gravity is addressed using the telegraph equation \citep{maeda2024improved}. 
Orbits of massive stars are calculated using the fourth-order Runge-Kutta method \citep[e.g.,][]{press1986numerical}.
The star formation process and the treatment of stellar feedback are described in detail in the next section.

\subsection{Treatment of Star Formation and Stellar Feedback}
\subsubsection{Star formation}
In this section, we describe the procedure for star formation. 

First, we examine whether each cell satisfies the criteria for star formation, consisting of the following three conditions \citep[c.f.,][]{fukushima2020star,hirai2021sirius}:
\begin{itemize}
    \item[1.] The density within the cell exceeds the density threshold $n_\mathrm{th}$. 

    \item[2.] The combined energy of thermal, kinetic, magnetic, and gravitational components is negative: $E_\mathrm{th}+E_\mathrm{kn}+E_\mathrm{mg}+E_\mathrm{gr}<0$. 

    \item[3.] The gravitational potential of the cell constitutes a local minimum. 
\end{itemize}
The density threshold is determined by assuming the density profile of the Larson–Penston solution as $n_{\mathrm{th}}=8.86 c_{\mathrm{s}}^{2} /\left(\pi G \Delta x^{2}\right)=2.56 \times 10^{4} \mathrm{~cm}^{-3}$ \citep{gong2012implementation}, where $\Delta x=0.195\ \mathrm{pc}$ and $c_\mathrm{s}$ is the sound speed of gas at $T=10~\mathrm{K}$.

The star-forming cells during a given time interval $\Delta t$ are stochastically selected by using the star formation efficiency (SFE) \citep[c.f.,][]{colin2013molecular, Fujii2015,hirai2021sirius}: 
\begin{equation}
\epsilon_{\mathrm{SFE}}=\epsilon_{\mathrm{SFE, ff}} \frac{\Delta t}{t_\mathrm{ff}} \label{eq:SFE_con}
\end{equation}
where $\epsilon_{\mathrm{SFE}}$ is the local star formation efficiency, $\epsilon_{\mathrm{SFE, ff}}$ is the star formation efficiency per free-fall time, $\Delta t$ is the time step of the simulation, and  $t_\mathrm{ff}$ is the free fall time for each cell. 
In this treatment, the SFE depends on the density, with high-density regions ($\sim 10^4~\mathrm{cm^{-3}}$) having an SFE consistent with the cluster-forming regions ($\gtrsim 20-30\%$) \citep[c..f.,][]{Fujii2015}.
During the calculations, $\epsilon_{\mathrm{SFE, ff}}$ is set to 0.02, which is consistent with the observed value \citep[e.g.,][]{krumholz2011universal}. 
By comparing the value of $\epsilon_{\mathrm{SFE}}$ to the random number ($\mathcal{R}$) from 0 to 1, we identify star-forming cells stochastically when $\epsilon_{\mathrm{SFE}}>\mathcal{R}$. 
This corresponds to identifying 2\% cells that satisfy the star formation conditions as star-forming cells during $1~t_\mathrm{ff}$.

After identifying the star-forming cells, some fraction of the gas is converted into star cluster particles. 
Specifically, half of the mass of the star-forming cells is replaced by the stars \citep[c.f.,][]{colin2013molecular}.  
Note that this operation results in $\epsilon_{\mathrm{SFE,ff}}=0.01$, effectively. 
To account for feedback effects from stars, it is crucial to know whether the resulting star cluster particles contain massive stars. We determine the stellar mass distribution within formed cluster particles using Kroupa's initial mass function (IMF) \citep{kroupa2001variation}:
\begin{equation} 
\begin{split}
&F_\mathrm{IMF} \propto M^{-\alpha},\\ 
&\alpha=
\begin{cases}
0.3,  & (0.01 \leq M / \mathrm{M}_{\odot}<0.08,)\\
1.3,  & (0.08 \leq M / \mathrm{M}_{\odot}<0.50,) \\
2.3,   & (0.50 \leq M /  \mathrm{M}_{\odot}<150,)
\end{cases}
\end{split}
\end{equation}
where $M$ is stellar mass.
We adopt a maximum stellar mass of $150\ \mathrm{M_\odot}$ following  \cite{fujii2022sirius}. The stellar mass is assigned by using the normalized cumulative IMF $G(M)$, with $M_\mathrm{star}=G^{-1}(\mathcal{R})$, where $\mathcal{R}$ represents a random number between 0 and 1. The cumulative IMF $G(M)$ is defined as:
\begin{equation}
    G(M)\equiv  \frac{\int^M_{0.01\ \mathrm{M_{\odot}}} F_\mathrm{IMF}\ dM }{\int_{0.01\ \mathrm{M_{\odot}}}^{150\ \mathrm{M_{\odot}}} F_\mathrm{IMF}\  dM .} \label{eq:GR}
\end{equation}
We repeat this procedure until the total assigned stellar mass matches the mass of the star cluster particles. It is worth noting that if the assigned massive star mass surpasses that of the formed star cluster particle, we subtract an equivalent mass of the formed massive star from the surrounding cells.

We then track the orbital evolution of the formed star cluster particles. Their initial position and velocity are determined by those of the star-forming cell.

\subsubsection{Stellar feedback}
We implement photoionization feedback from star cluster particles containing massive stars ($>10\ \mathrm{M_{\odot}}$). This feedback is introduced as heating within the Strömgren radius, $R_\mathrm{s}$. At this radius, the ionization rate from the massive star balances the recombination rate of the ionized gas, given by:
\begin{equation}
    R_{\mathrm{s}} \equiv\left(\frac{3}{4 \pi} \frac{S_{*}}{\alpha_\mathrm{B} n^{2}}\right)^{1 / 3},
\end{equation} 
where $\alpha_\mathrm{B}=2.6\times10^{-13}\ \mathrm{cm^3\ s^{-1}}$ represents the recombination coefficient \citep{2006agna.book.....O} and $S_{*}$ the ionizing photons emitted by massive stars per unit time. 
Since the gas structure around massive stars is inhomogeneous, we estimate the Strömgren radius by using the recombination rate equation:
\begin{equation}
    U(r) =\int_0^{r} \alpha_\mathrm{B} n^{2} \cdot 4\pi r^2\ dr, 
\end{equation}
where $r$ is the distance from a star cluster particle. 
Here, $r$ corresponds to the Strömgren radius when $S_{*}=U(r)$ is satisfied. 
Within the Strömgren radius, the gas is heated up to $10^4\ \mathrm{K}$, which is the typical temperature of HII regions \citep[e.g.,][]{fujii2022sirius,colin2013molecular}. 
The effect of photoionization heating is considered as the term $S_\mathrm{MS}$ in eq.~(\ref{eq:engy-feed}). 
We calculate the ionizing photons from star cluster particles as the sum of the photons from the massive stars they contain ($S_* = \sum_{i} S_{*,i}$). 
Following \cite{fujii2022sirius}, we calculate $S_{*,i}$, ionizing photon emissivity from $i$-th massive star, as a function of stellar mass, derived by fitting OSTAR2002 data \citep{lanz2003grid}: 
\begin{equation}
    \log (S_{*,i})=a+b x+c x^{2}+d x^{3}+e x^{4}+f x^{5},\ (x=\log M_i).
\end{equation}
The best fit parameters are: $a=-39.3178$, $b=221.997$, $c=-227.456$, $d=117.410$, $e=-30.1511$, and $f=3.06810$. 

\subsection{Simulation setup}
Here, we briefly summarize the setup of the fiducial simulation model in \cite{maeda2021formation}, to which we add the feedback effect.

As described above, we resolve the $100^3~\mathrm{pc^3}$ box in uniform cells of $512^3$ (cell size $\sim 0.2~\mathrm{pc}$). 
The direction of the colliding flow is parallel to the $x$-axis.
Periodic boundary conditions are applied at the $y, z$ boundaries, while continuous gas flow boundary conditions are imposed at the $x$ boundaries. The gravitational potential at the $x$ boundaries is computed using the method developed by \cite{miyama1987fragmentation}.

We consider a fiducial mean initial density for the HI gas, $\left\langle n_{0}\right\rangle \sim 1~\mathrm{cm^{-3}}$, which represents the typical value of the HI gas density. However, to assess the effects of density variations, we also investigate cases with an average density of $10~\mathrm{cm^{-3}}$. 
We adopt a collision velocity of $v_0=100~\mathrm{km\ s^{-1}}$, typical for collisions observed between HI gas flows and the disk in the LMC.
Additionally, to confirm the importance of high-velocity gas collisions, simulations are conducted for the case of slow-velocity collisions ($v_0=15\ \mathrm{km\ s^{-1}}$). A summary of the initial parameters is provided in Table \ref{tab:models}.

We introduce density fluctuations using Kolmogorov's power spectrum, where the density dispersion is set to $20\%$ of the average density $\left\langle n_{0}\right\rangle$.
The initial thermal pressures are set to $4.5\times10^3$ $(1.6\times10^3)$ $\mathrm{K\ cm^{-3}}$ for cases where the average density $\left\langle n_{0}\right\rangle$ is approximately $1$ $(10)~\mathrm{cm^{-3}}$, corresponding to temperatures of $T= 4.5\times 10^3$ ($1.6\times 10^2$, respectively) $\mathrm{K}$.

A uniform magnetic field, oriented at $45^{\circ}$ relative to the $x$-axis in the $x$-$y$ plane, is initially set with a strength of $1.0\ \mathrm{\mu G}$, consistent with LMC observations using the rotation measure \citep{gaensler2005magnetic}.
For the case of dense gas ($\left\langle n_{0}\right\rangle \sim 10~\mathrm{cm^{-3}}$), we use $3.0\ \mathrm{\mu G}$ as described in \cite{maeda2021formation}. 

The gas metallicity is set to the solar value, which is consistent with the values observed in YMC forming regions \citep{portegies2010young} and the initial chemical abundances are  
$x_{\mathrm{H}} \equiv n_{\mathrm{H}} / \sum n_{\mathrm{i}}=0.91$, $x_{\mathrm{p}} = 9.4 \times 10^{-3}$, $x_{\mathrm{H}_{2}}=9.4 \times 10^{-9}$, $x_{\mathrm{He}}=9.0 \times 10^{-2}$, $x_{\mathrm{He}^{+}}=5.6 \times 10^{-4}$, $x_{\mathrm{C}^{+}}=1.4 \times 10^{-4}$, $x_{\mathrm{C}}=1.6 \times 10^{-9}$, $x_{\mathrm{CO}}=2.2\times 10^{-21}$, $x_{\mathrm{O}}=3.2 \times 10^{-4}$, and the background far ultraviolet (FUV) intensity is set to the Habing flux, $1.6 \times 10^{-3} \ \mathrm{erg}\ \mathrm{cm}^{-2}\ \mathrm{s}^{-1}$ \citep{habing1968interstellar}. 

\begin{deluxetable*}{lccc}
\tablenum{1}
\tablecaption{Parameters of our simulations\label{tab:models}}
\tablewidth{0pt}
\tablehead{
\colhead{Model} &\colhead{$v_\mathrm{col}$ ($\mathrm{km~s^{-1}}$)}& \colhead{ $\left\langle n_{0}\right\rangle$ ($\mathrm{cm^{-3}}$)} & \colhead{$B_{0}$ ($\mathrm{\mu G}$)} }
\startdata
V100D1      &   $100 $   & $ 1.0 $    &   $1.0$       \\
V100D10      &   $100 $   & $ 10.0$    &   $3.0$       \\
V15D1          &   $15 $  & $ 1.0 $    &   $1.0$     \\
V15D10         &   $15 $   & $ 10.0$    &   $3.0$    \\
\enddata
\end{deluxetable*}

\section{Results}
Our primary focus is on whether YMC formation is possible through fast collisions. First, in Sec.~\ref{sec:fast}, we examine the evolution of post-shock gas and investigate the properties of the clumps formed therein, as well as the potential for their evolution into YMCs, in the case of fast collisions ($v_0=100\ \mathrm{km\ s^{-1}}$). Subsequently, in Sec.~\ref{sec:slow}, we explore how these outcomes change in the case of slower collisions ($v_0=15\ \mathrm{km\ s^{-1}}$).

\subsection{Fast ($v_0=100~\mathrm{km~s^{-1}}$) gas collisions}\label{sec:fast}
Here, we first describe in detail the fiducial case with an initial mean density of $n_0=1\ \mathrm{cm^{-3}}$. Similar to our previous work \citep{maeda2021formation}, we confirm the formation of massive clumps that could serve as progenitors to YMCs. We observe that relatively small clumps ($\lesssim 10^4~\mathrm{M_{\odot}}$) are strongly affected by stellar feedback, disturbing their evolution into massive star clusters while larger clumps ($\gtrsim 10^5~\mathrm{M_{\odot}}$) can overcome stellar feedback due to strong gravity, enabling their evolution into YMCs.

We also see the case with a higher density ($n_0=10\ \mathrm{cm^{-3}}$), where we observe the formation of even larger clumps due to the following reason.
From observation of the Antennae galaxies, larger density gas collisions than the fiducial model are suggested around very massive YMCs ($\gtrsim10^{6}~\mathrm{M_\odot}$) \citep{tsuge2021formation}. 
The derived density and velocity of colliding gas is $\gtrsim10-100~\mathrm{cm^{-3}}$ with velocity of colliding gas $\sim100~\mathrm{km~s^{-1}}$. 

In this section, we show the results of the high-density gas collision with $100~\mathrm{km~s^{-1}}$ to find out the relationship between the ram pressure (the density of colliding flow) and YMC mass.

\subsubsection{Fiducial density ($n_{0}=1{\rm cm^{-3}}$) case (V100D1)}
\begin{figure*}[!t]
\centerline{
\includegraphics[clip, width=0.8\textwidth]{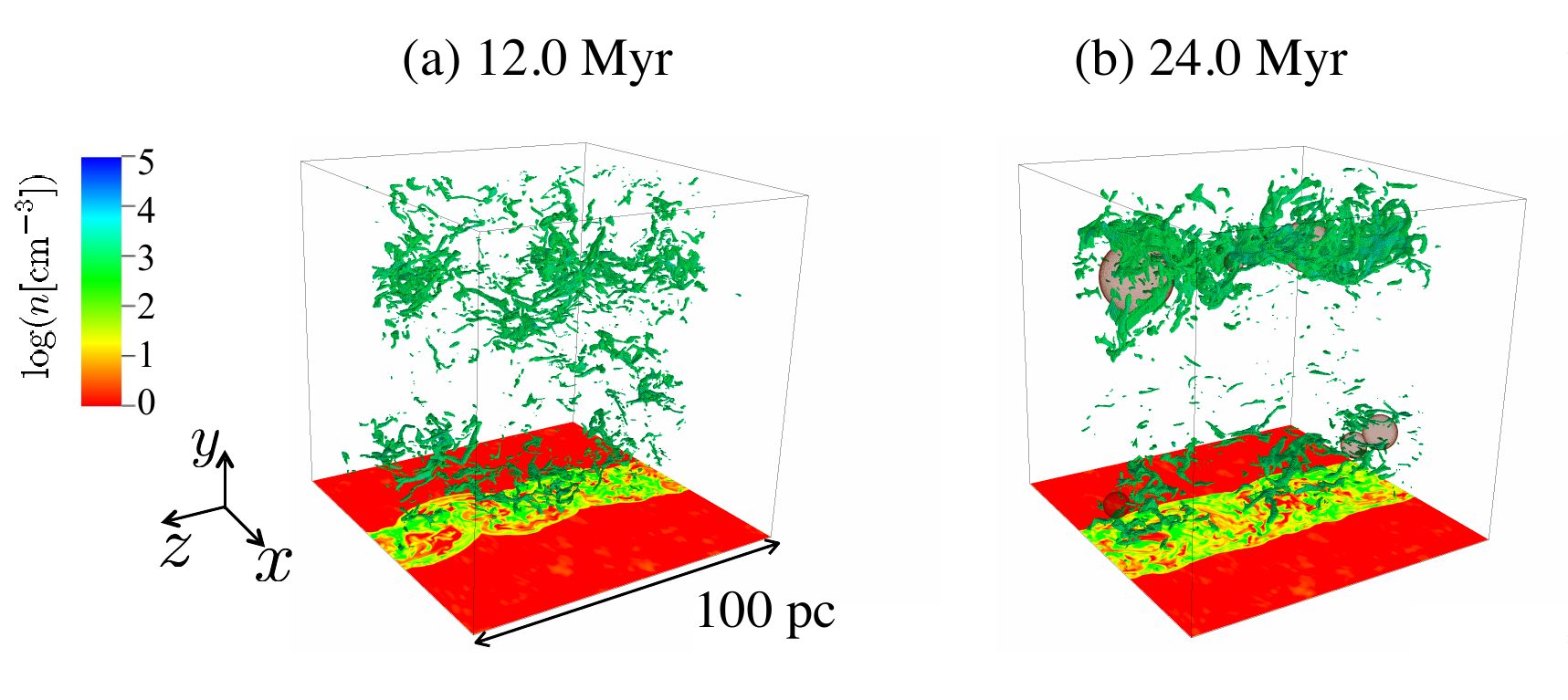}}
\caption{Snapshots of the three-dimensional structure of high-density gas $(n>10^3\ \mathrm{cm^{-3}})$ at two distinct epochs, (a) $t=12.0$ and (b) $t=24.0$ Myr. The red spheres in the box illustrate formed HII regions. The bottom panels display two-dimensional density cross-sections on the $x-z$ plane at $y=0$.
\label{fig:d125250}}
\end{figure*}

We first investigate the consequences of the fast ($v_0=100\ \mathrm{km\ s^{-1}}$) collision between HI gas of typical ISM density $n_0=1\ \mathrm{cm^{-3}}$. 
To this end, 
we present the snapshots of three-dimensional density fields at (a) $t=12.0~\mathrm{Myr}$, and (b) $24.0~\mathrm{Myr}$ in Figure \ref{fig:d125250}. 
The boxes represent the entire computational domain, and the panels at the bottom of the boxes show a two-dimensional density cross-section at $y=0$. 
In the box, we only illustrate high-density regions ($n>10^3\ \mathrm{cm^{-3}}$; green) and HII regions (red spheres). 

The collision of gas generates two shock waves propagating outwards in the $+x$ and $-x$ directions, respectively (see the bottom panels of Figure \ref{fig:d125250}). 
After the cooling timescale ($\sim 1~\mathrm{Myr}$), thermal pressure is removed from the postshock layer, and the postshock gas starts to contract.  
The contraction terminates once the magnetic pressure at the postshock region balances with the ram pressure of the converging flows \citep{inoue2009two}. 
The averaged density of the postshock layer at this epoch is given by
\begin{equation}
\bar{n} \sim 35\ \mathrm{cm}^{-3}\left(\frac{v_{\mathrm{sh}}}{50\ \mathrm{km}\ \mathrm{s}^{-1}}\right)\left(\frac{n_{0}}{1\   \mathrm{cm}^{-3}}\right)^{\frac{3}{2}}\left(\frac{B_{0,\perp}}{1\ \mu \mathrm{G}}\right)^{-1},  \label{eq:shockjump-n}
\end{equation}
where $n_{\mathrm{0}}$ represents the number density of the preshock region, $B_{\mathrm{0},\perp}$ component of the magnetic field perpendicular to the converging flows,
and $v_{\mathrm{sh}}$ the shock velocity \citep{inoue2009two,maeda2021formation}. 
Because the postshock HI gas is thermally unstable, the gas evolves into two components: a low-temperature, high-density gas (cold neutral medium; CNM) and a high-temperature, low-density gas (warm neutral medium; WNM). 
The two-phase gas structure as a consequence of the thermal instability is observed in Figure \ref{fig:d125250}.

The free-fall time of the shocked sheet can be estimated as 
\begin{equation}
    t_{\mathrm{ff, sheet}}=\frac{1}{\sqrt{2 \pi G \bar{\rho}}}\sim 6\ \mathrm{Myr}\left(\frac{\bar{n}}{35\ \mathrm{cm}^{-3}}\right)^{-\frac{1}{2}}, \label{eq:tff}
\end{equation}
where $\bar{\rho}$ is the average shocked gas density. 
After a few $t_\mathrm{ff, sheet}$, the CNMs on the postshock region are collected by the effect of gravity to form more massive gas clumps \citep{maeda2021formation} (Figure \ref{fig:d125250}b). 
When a sufficiently massive gas clump forms, 
stars including massive stars are created in these gas clumps. 
Once a massive star forms, it creates an HII region (Figure \ref{fig:d125250}b) and heats up the surrounding gas. 

\begin{figure}[htbp]
\centering
\includegraphics[width=0.4\textwidth]{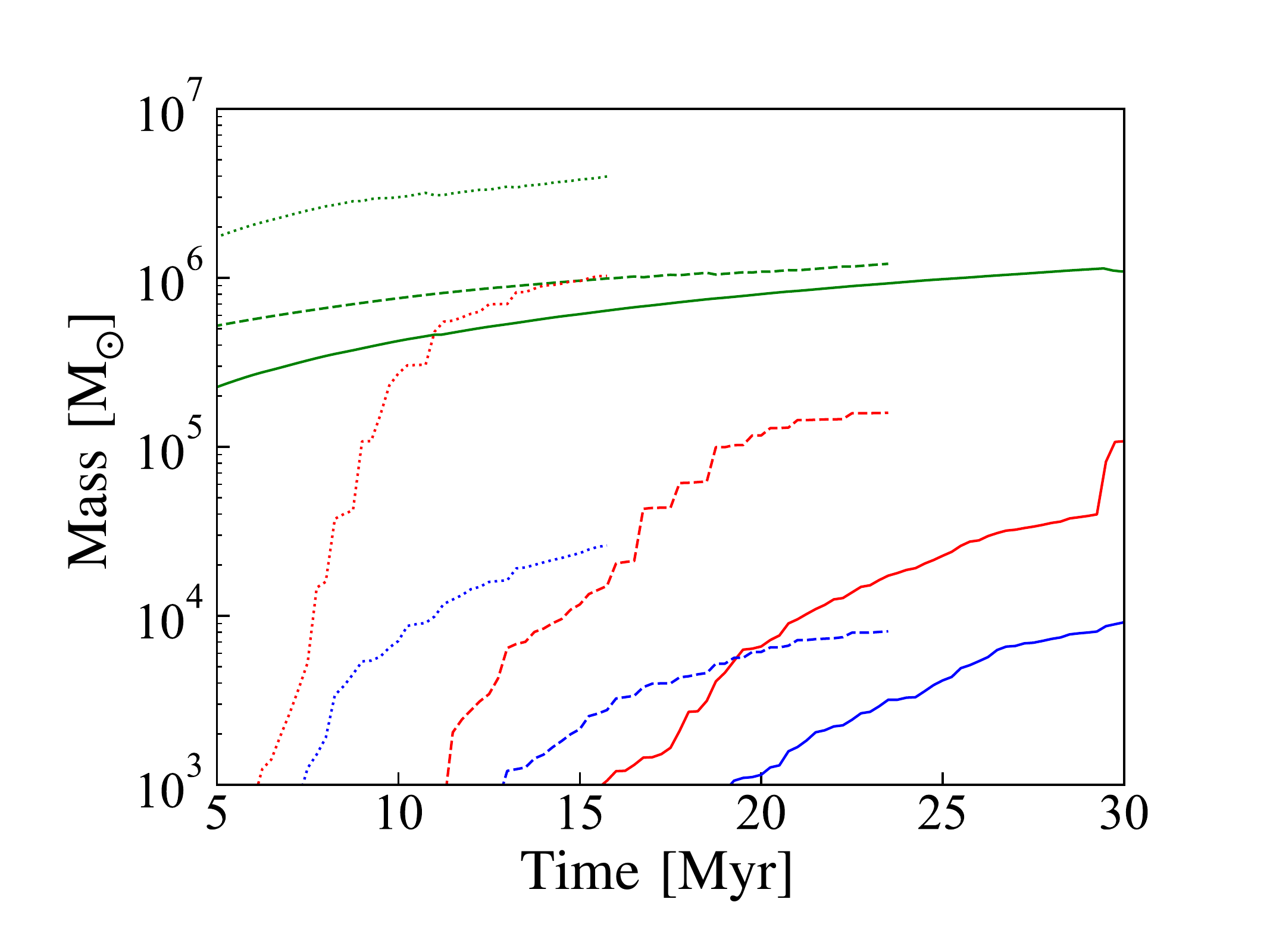}
\caption{The evolution of gas mass (green), stellar mass (red), and mass of massive stars (blue) across the entire simulation box. Results for the fiducial case (V100D1) are presented as solid lines, adding that those for other cases are also displayed (V100D10: dotted, V15D10: dashed). 
\label{fig:TotalMass}}
\end{figure}

\begin{figure}[htbp]
\centering
\includegraphics[width=0.4\textwidth]{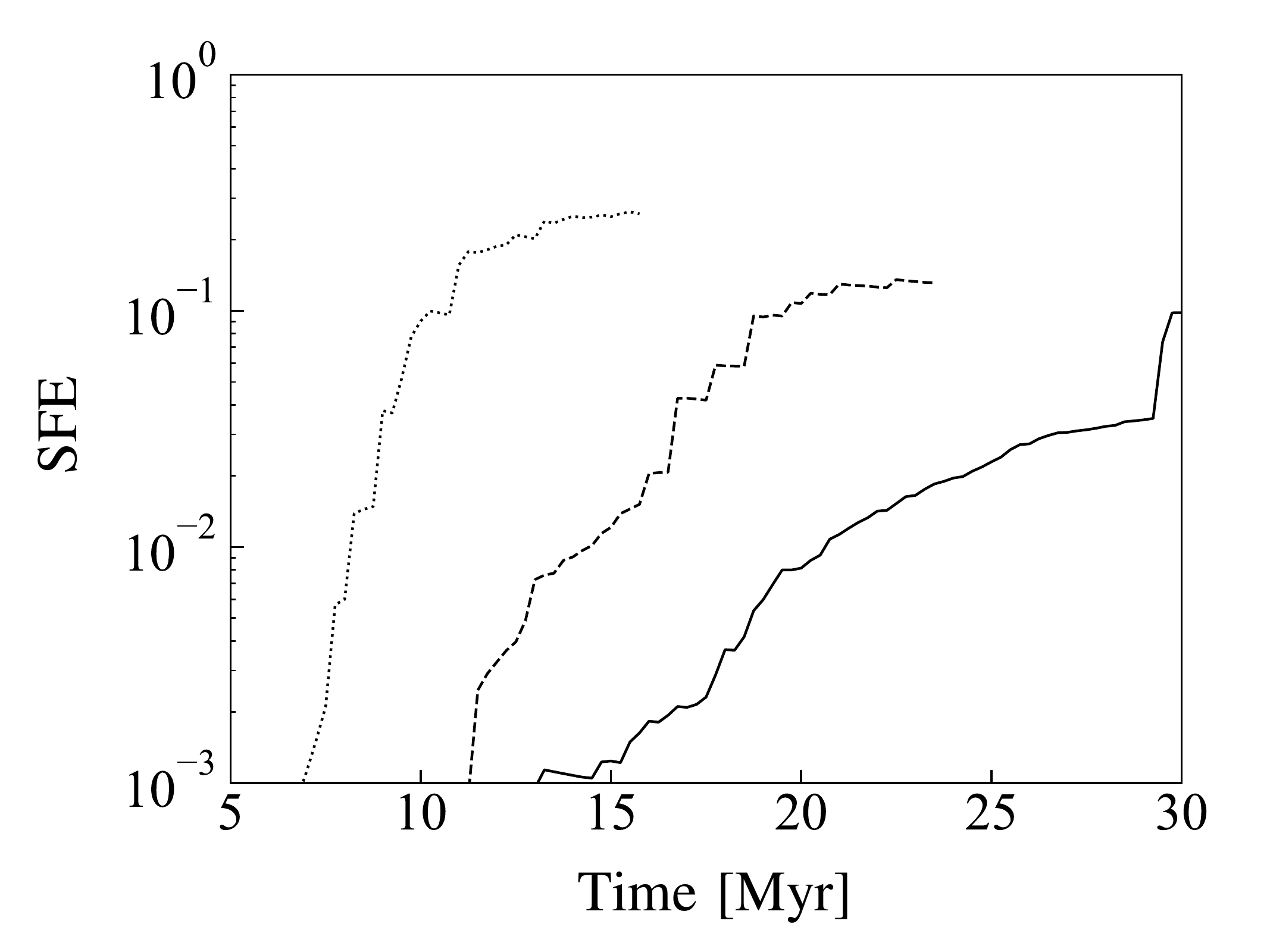}
\caption{The evolution of star formation efficiency (SFE) across the entire gas in the simulation box. 
The solid, dashed, and dotted lines show the results of models V100D1, V15D10, and V100D10, respectively, as in Figure \ref{fig:TotalMass}. 
\label{fig:totSFE}}
\end{figure}

The evolution of the total gas mass (green), stellar mass (red), and mass of massive stars (blue) within the simulation domain is illustrated in Figure \ref{fig:TotalMass}. Here, alongside the fiducial case (V100D1, solid line), we also present other cases, which will be described later (dotted for V100D10 and dashed for V15D10) for comparison. 
In the late stage of the simulations, star formation occurs in all models. 

\begin{figure}[htbp]
\centering
\includegraphics[width=0.4\textwidth]{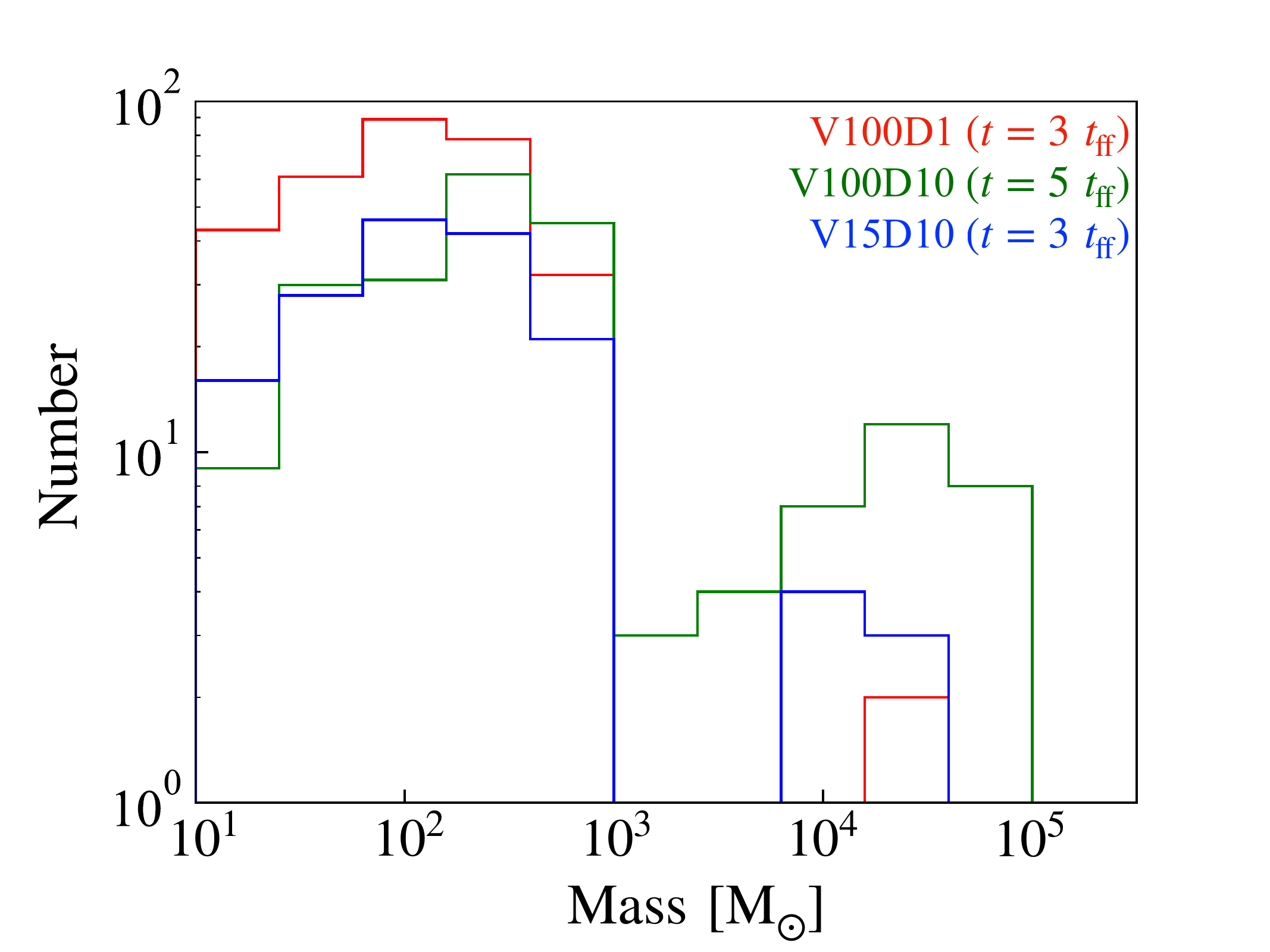}
\caption{The mass distributions of formed star cluster particles in runs V100D1 (red), V100D10 (green), and V15D10 (blue). 
\label{fig:particl}}
\end{figure}

The total star formation efficiency (SFE), calculated as the total stellar mass divided by the total gas mass within the entire simulation volume, is depicted in Figure \ref{fig:totSFE}. 
Here, the SFE depends on the density as described by eq.~(\ref{eq:SFE_con}), and the values shown in Figure \ref{fig:totSFE} represent the average over the entire numerical domain.
The solid, dotted, and dashed lines again indicate the results of models V100D1, V100D10, and V15D10, respectively. 
Note that the final SFE in the fiducial model (V100D1) reaches approximately $10\%$, consistent with observations of regions known for producing YMCs \citep[e.g.,][]{fukui2017formation}. 
The mass of the formed star cluster particles is shown in Figure \ref{fig:particl}. From the red line in the figure, it is evident that star cluster particles with typical YMC masses of $>10^4~\mathrm{M_\odot}$ are formed. Note, however, that star cluster particles have no spatial extent, unlike realistic YMCs. For this reason, we focus on the gas clumps, which are considered to be the precursors to YMCs, rather than the cluster particles themselves in the following sections.

\begin{figure}[htbp]
\centering 
\includegraphics[width=0.4\textwidth]{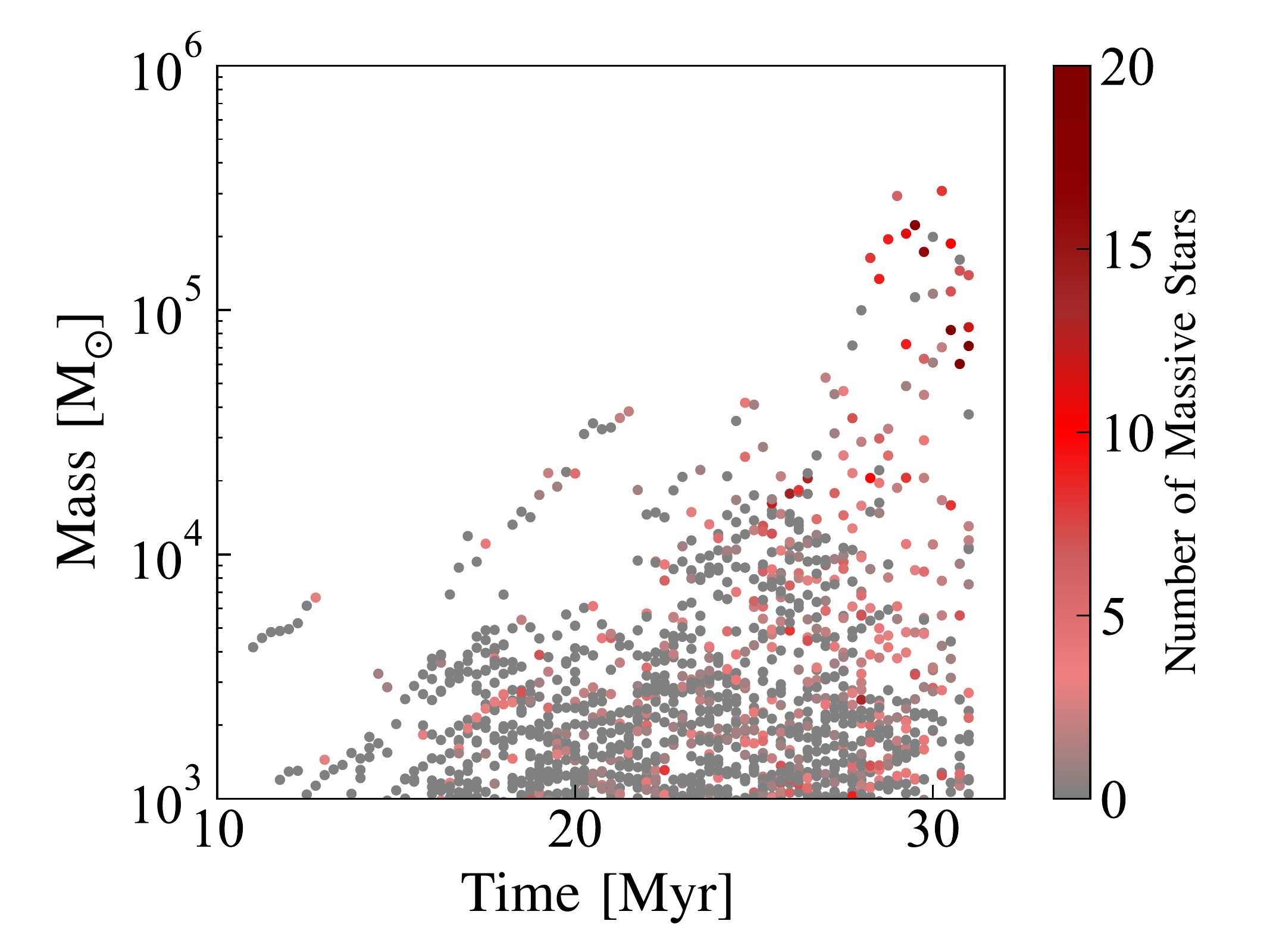}
\caption{The mass of clumps identified at each time step in the fiducial run (V100D1). Each data point represents an individual clump, with the color indicating the number of massive stars surrounding the clump.
\label{fig:clmp-HII}}
\end{figure}

To investigate the candidates of the YMC-forming clumps, which are expected to be massive $>10^4\ \mathrm{M_{\odot}}$ and compact $\sim\mathrm{pc}$, we identify dense gas clumps by using the Friends-of-Friends algorithm at each time. 
Here, we define dense gas clouds as the connected region with a number density larger than $n=10^4\ \mathrm{cm^{-3}}$ following \cite{maeda2021formation}. 
Figure \ref{fig:clmp-HII} illustrates the evolution of the identified dense clump masses, where each data point represents the mass of an identified clump at a given time. 
This figure clearly shows an increase in clump mass over time: clumps exceeding $10^4~\mathrm{M_\odot}$ begin to form at $t=17~\mathrm{Myr}$, and by $t=27~\mathrm{Myr}$, the most massive gas clump has reached a mass of $10^5~\mathrm{M_\odot}$. 
The most massive clump formed in this simulation attains a size of $\sim 5\ \mathrm{pc}$ and a mass of $\sim 1.4\times 10^5\ \mathrm{M_\odot}$ at $t=30~\mathrm{Myr}$, corresponding to a surface density of $\Sigma \sim 7\times10^3~\mathrm{M_\odot}~\mathrm{pc^{-2}}$. 
On the other hand, the trajectory of some clumps (mostly those with $M < 10^4~\mathrm{M_\odot}$) seem to be destroyed and lose their mass at a later time. 

Within these massive clumps, the formation of massive stars takes place.  
The color in Figure \ref{fig:clmp-HII} indicates the number of massive stars around the clump. 
Specifically, this count includes massive stars within a radius of $2.5~\mathrm{pc}$ from the clump's center of gravity. 
Gray dots represent clumps that have not formed massive stars, and red dots represent clumps having more than a dozen massive stars. 
Note that, around the breaking point, $(t, M)\simeq (12~\mathrm{Myr}, 7\times10^{3}~\mathrm{M_\odot})$ for example, in the evolutionary trajectory of clumps with masses $\lesssim 10^4~\mathrm{M_\odot}$, we observe the formation of massive stars, which are subsequently influenced by feedback processes. 
Conversely, the figure also illustrates that even in the presence of feedback, the most massive clump (with masses on the order of $\sim10^5~\mathrm{M_\odot}$) can persist for extended periods.

\begin{figure}[htbp]
\centering
\includegraphics[width=0.4\textwidth]{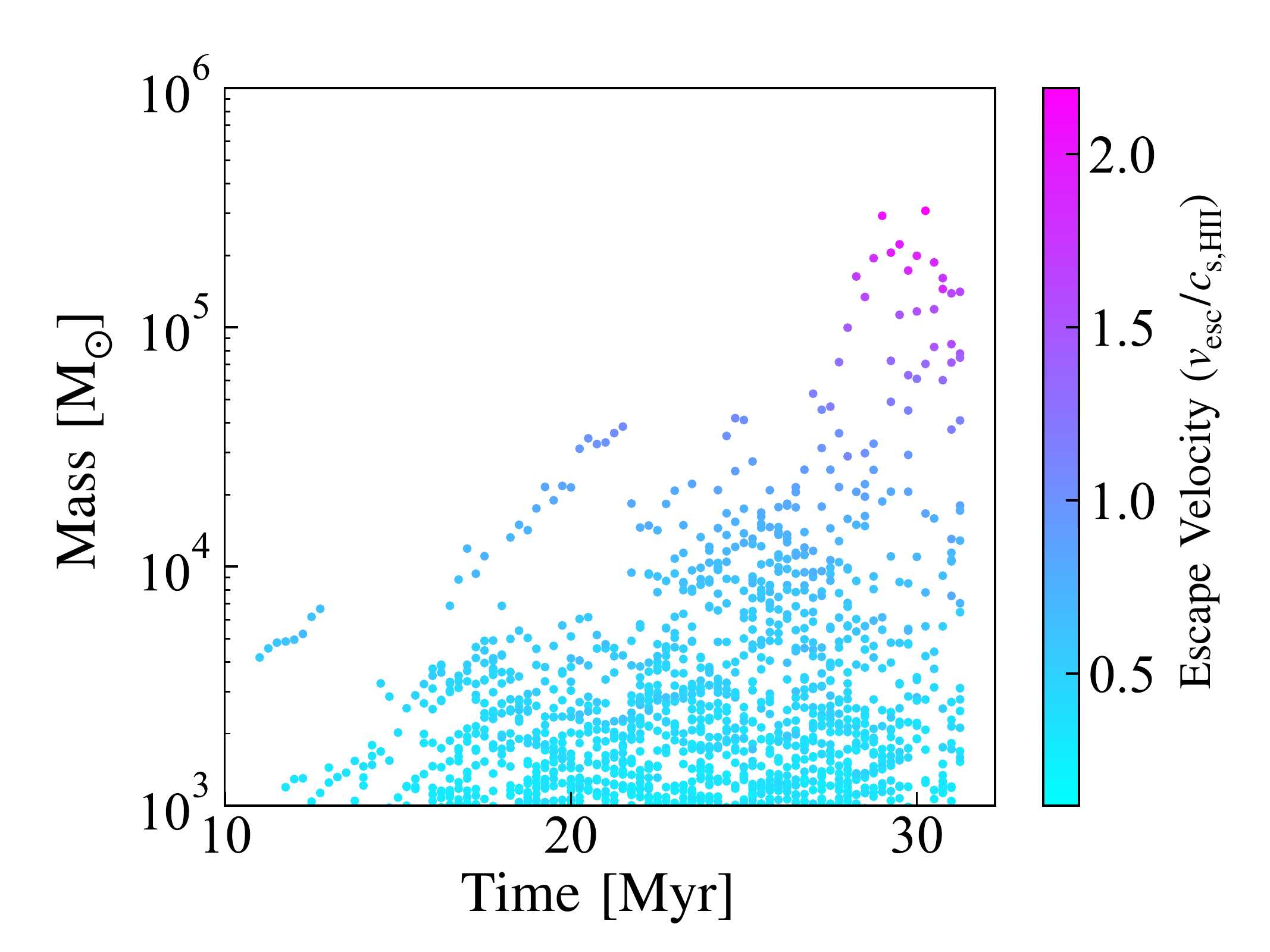}
\caption{The same as Figure \ref{fig:clmp-HII}, but the color of each data point indicates the escape velocity of each clump. 
\label{fig:clmp-vesc}}
\end{figure}

To assess the influence of the feedback for different mass clumps, we examined the average escape velocity of the clumps (see Figure \ref{fig:clmp-vesc}). 
The escape velocity is determined as follows: 
\begin{equation}
v_{\mathrm{esc}} = \sqrt{\frac{2 M_\mathrm{cl} G}{r_\mathrm{cl}}},
\end{equation}
where 
$M_\mathrm{cl}$ is the clump mass, and $r_\mathrm{cl}$
its radius, approximated as the cubic root of its volume.

Figure \ref{fig:clmp-vesc} depicts the mass evolution of the clumps with its escape velocity, represented by color. 
Here, the escape velocity is normalized against the sound speed within the HII region ($c_\mathrm{s,HII}\sim 10~\mathrm{km~s^{-1}}$). 
The massive gas clumps with $\sim 10^5\ \mathrm{M_\odot}$ have an escape velocity larger than $c_\mathrm{s,HII}$, indicating that they survive against the feedback thanks to their compactness.  
Conversely, for clumps with masses around $\lesssim 10^4\ \mathrm{M_\odot}$, their escape velocities are at most the sound speed of the HII region. 
These clumps cannot bind themselves by gravity and eventually evaporate by the feedback. 

\begin{figure*}[!t]
\centerline{
\includegraphics[clip, width=0.8\textwidth]{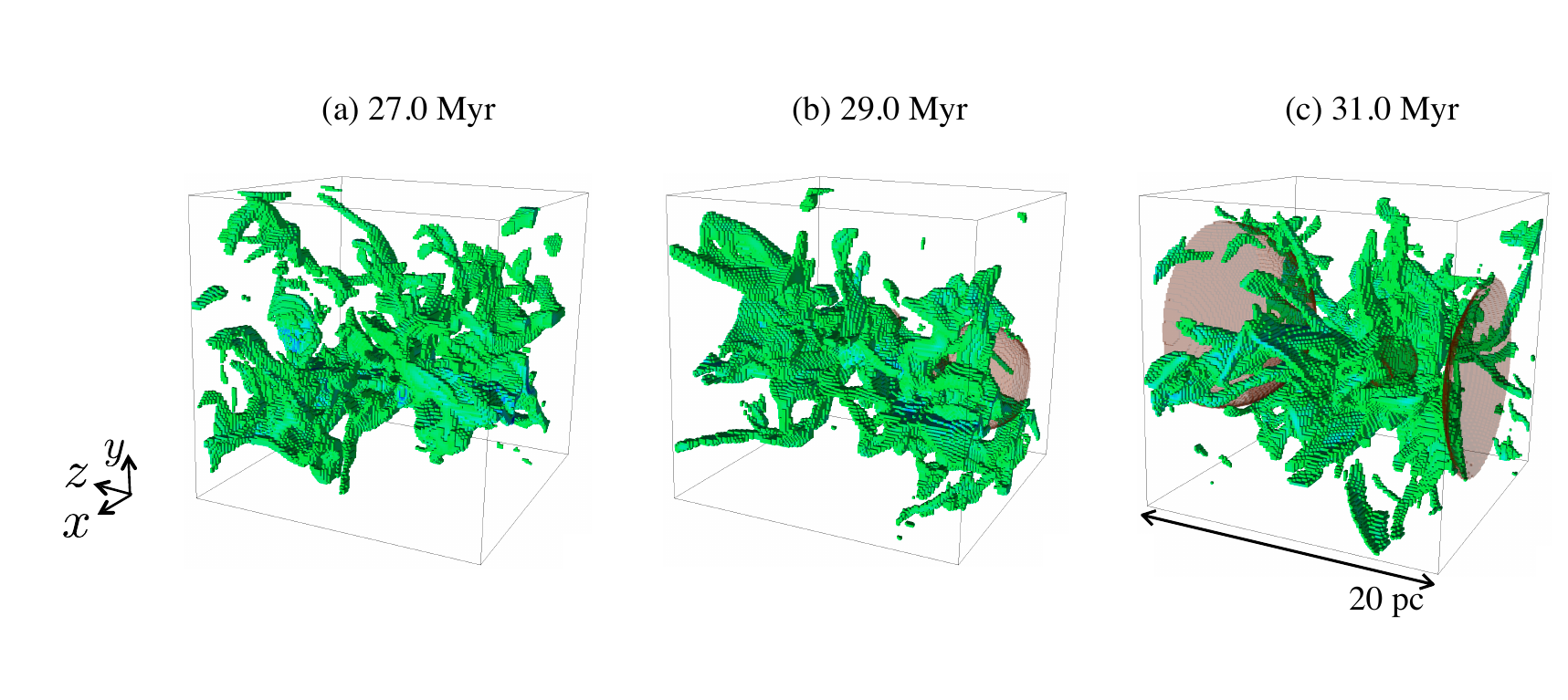}}
\caption{
The detailed view of the density structure surrounding the most massive gas clump, within a cube with $20~\mathrm{pc}$ sides. High-density regions $(n>10^3\ \mathrm{cm^{-3}})$ are plotted for three different time steps, $t=\mathrm{(a)}~27.0$, $\mathrm{(b)}~29.0$, and $\mathrm{(c)}~31.0~\mathrm{Myr}$. The red regions indicate forming HII regions. Despite the development of the HII regions, the clump retains its gas as its escape velocity exceeds the sound speed of the ionized gas.
\label{fig:reg1-feed}}
\end{figure*}

\begin{figure*}[!t]
\centerline{
\includegraphics[clip, width=0.8\textwidth]{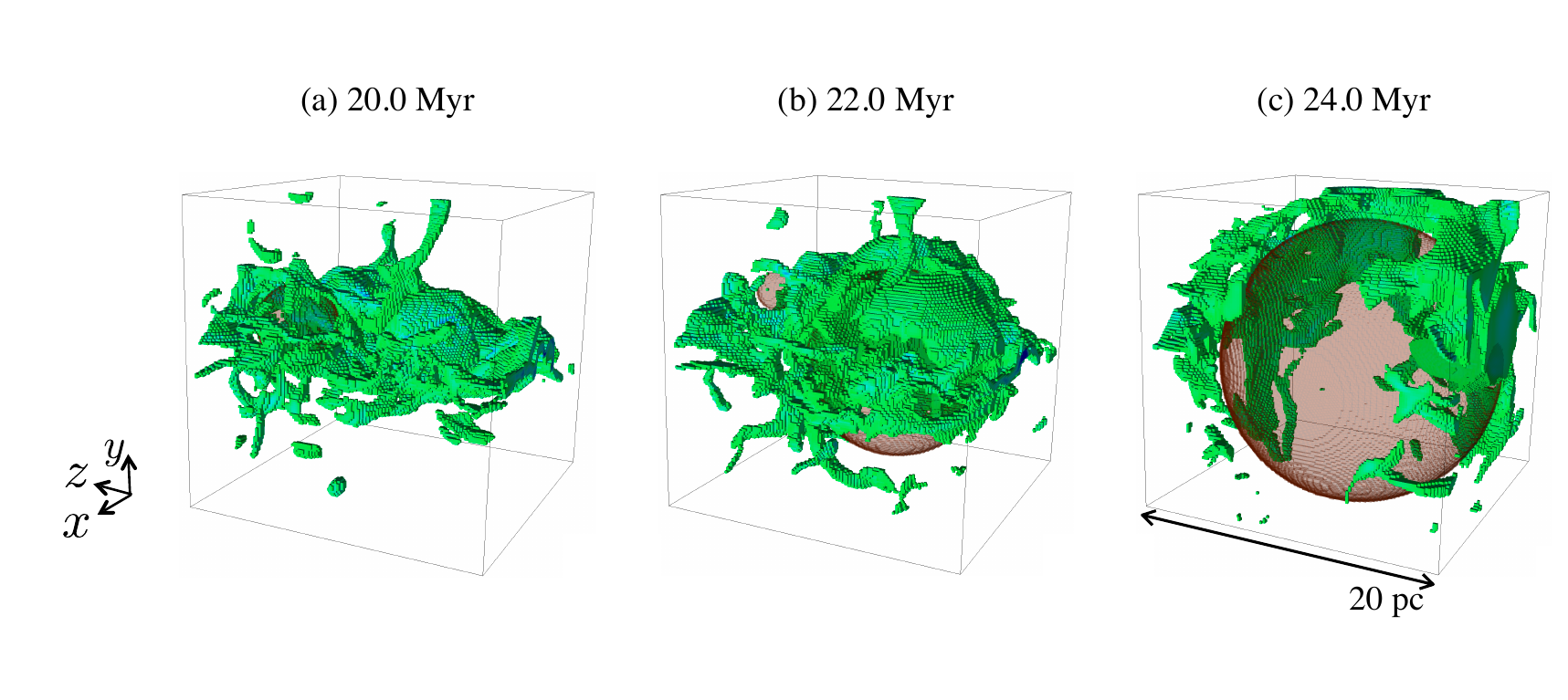}}
\caption{
Similar to depicted in Figure \ref{fig:reg1-feed}, but focusing on a less massive gas clump ($\sim 3\times 10^4\ \mathrm{M_\odot}$). Density structures within a cube with $20~\mathrm{pc}$ sides are illustrated for three snapshots, $t=\mathrm{(a)}~20.0$, $\mathrm{(b)}~22.0$, and $\mathrm{(c)}~24.0~\mathrm{Myr}$. Due to the clump's escape velocity being smaller than the sound speed of the ionized gas, the growth of the HII region results in the disruption of the clump. 
\label{fig:reg2-feed}}
\end{figure*}

For a more detailed examination of this phenomenon, we present the density structure of the surrounding dense gas ($>10^3\ \mathrm{cm^{-3}}$) for the most massive gas clump ($\sim 1 \times 10^5\ \mathrm{M_\odot}$ with $v_\mathrm{esc}\sim19.0~\mathrm{km~s^{-1}}$) and a gas clump with a mass of approximately $3\times 10^4\ \mathrm{M_\odot}$ ($v_\mathrm{esc}\sim9.4~\mathrm{km~s^{-1}}$) in Figures \ref{fig:reg1-feed} and \ref{fig:reg2-feed}, respectively. 
Panels (a) to (c) represent the time evolution in the figures. Additionally, the red regions indicate the HII regions. Around the massive clump where the escape velocity exceeds the ionized gas sound speed (Figure \ref{fig:reg1-feed}), despite the formation of the HII region, the entire cloud remains mostly intact due to gravity. 
Conversely, around the smaller clump where the escape velocity is below the ionized gas sound speed, expansion of the HII region destructs the clump (Figure \ref{fig:reg2-feed}).

\begin{figure*}[!t]
\centerline{
\includegraphics[clip, width=1.\textwidth]{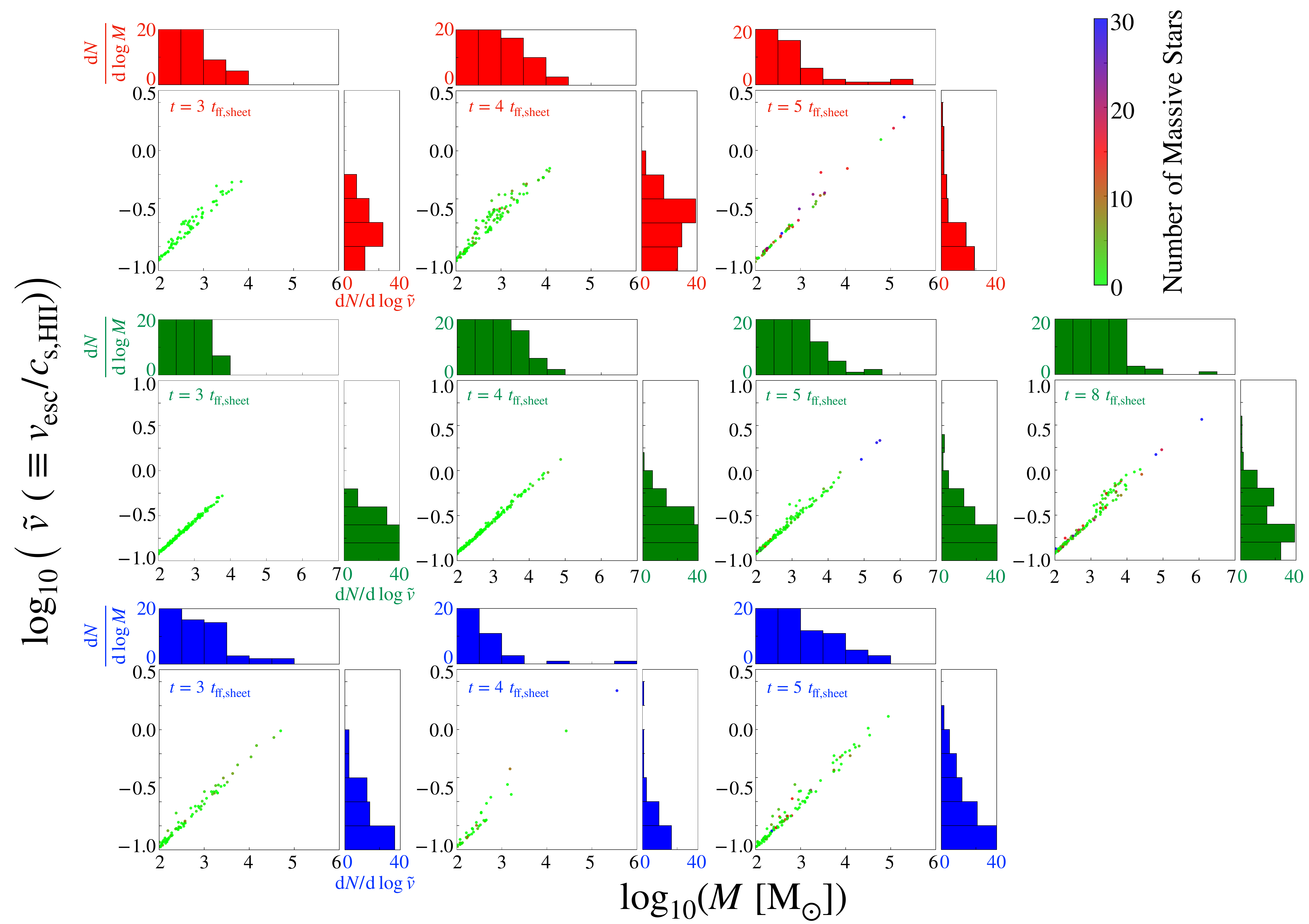}}
\caption{
Scatter plots of the mass and escape velocity for the gas clumps formed in each model. Models V100D1, V100D10, and V15D10 are represented from top to bottom. Left to right indicates the results at different time steps $t=3, 4, 5,$ and $8~t_\mathrm{ff,sheet}$. The color of the plots means the number of massive stars around the clumps. The upper and left sides of each panel are attached to the mass and the escape velocity distribution functions, respectively. Note that the middle panels (V100D10 model) have different scales on the vertical and horizontal axes than the others. 
\label{fig:hist_sct}}
\end{figure*}

Figure \ref{fig:hist_sct} shows the scatter plot of the clump mass and escape velocity, with the colors of the dots corresponding to the number of massive stars around ($<2.5~\mathrm{pc}$) the clump as Figure \ref{fig:clmp-HII}. 
The results of the different models are displayed from top to bottom (top; V100D1, middle; V100D10, and bottom; V15D10), and from left to right, the results for different time steps, $t=3, 4, 5,$ and $8~t_\mathrm{ff, sheet}$. 
Distribution functions for mass and escape velocity are also shown in each panel, corresponding to model V100D1 in red, model V100D10 in green, and model V15D10 in blue. 
In the fiducial model (red), the clump mass function at $t=18\ \mathrm{Myr}$, $t=24\ \mathrm{Myr}$, and $t=30\ \mathrm{Myr}$ are shown in Figure \ref{fig:hist_sct}.  
This figure reveals a notable trend wherein the number of clumps with masses around $\sim 10^4\ \mathrm{M_\odot}$ experiences a decrease with time, attributed to the influence of feedback mechanisms. On the other hand, it also demonstrates that the massive gas clumps ($\sim 10^5\ \mathrm{M_\odot}$) maintain their substantial mass over prolonged periods, even containing many massive stars. 

These simulations show that even under the presence of feedback, massive ($> 10^5\ \mathrm{M_\odot}$) and compact ($\sim 5\mathrm{pc}$) gas clumps formed by the fast HI gas collision can survive.  
The formed massive gas clumps are so compact that their escape velocity is greater than the sound speed of the HII region. In such cases, stars are formed with a high SFE ($>20-30\%$), and the clump can evolve into a YMC. Specifically, to form a massive star cluster, the clump needs to have a mass of $>10^5\ \mathrm{M_\odot}$ and a surface density of $>200\ \mathrm{M}_{\odot}\ \mathrm{pc}^{-2}$ \citep{fukushima2022far}.

\subsubsection{Dense gas collisions (V100D10)}
\begin{figure}[htbp]
\centering
\includegraphics[width=0.4\textwidth]{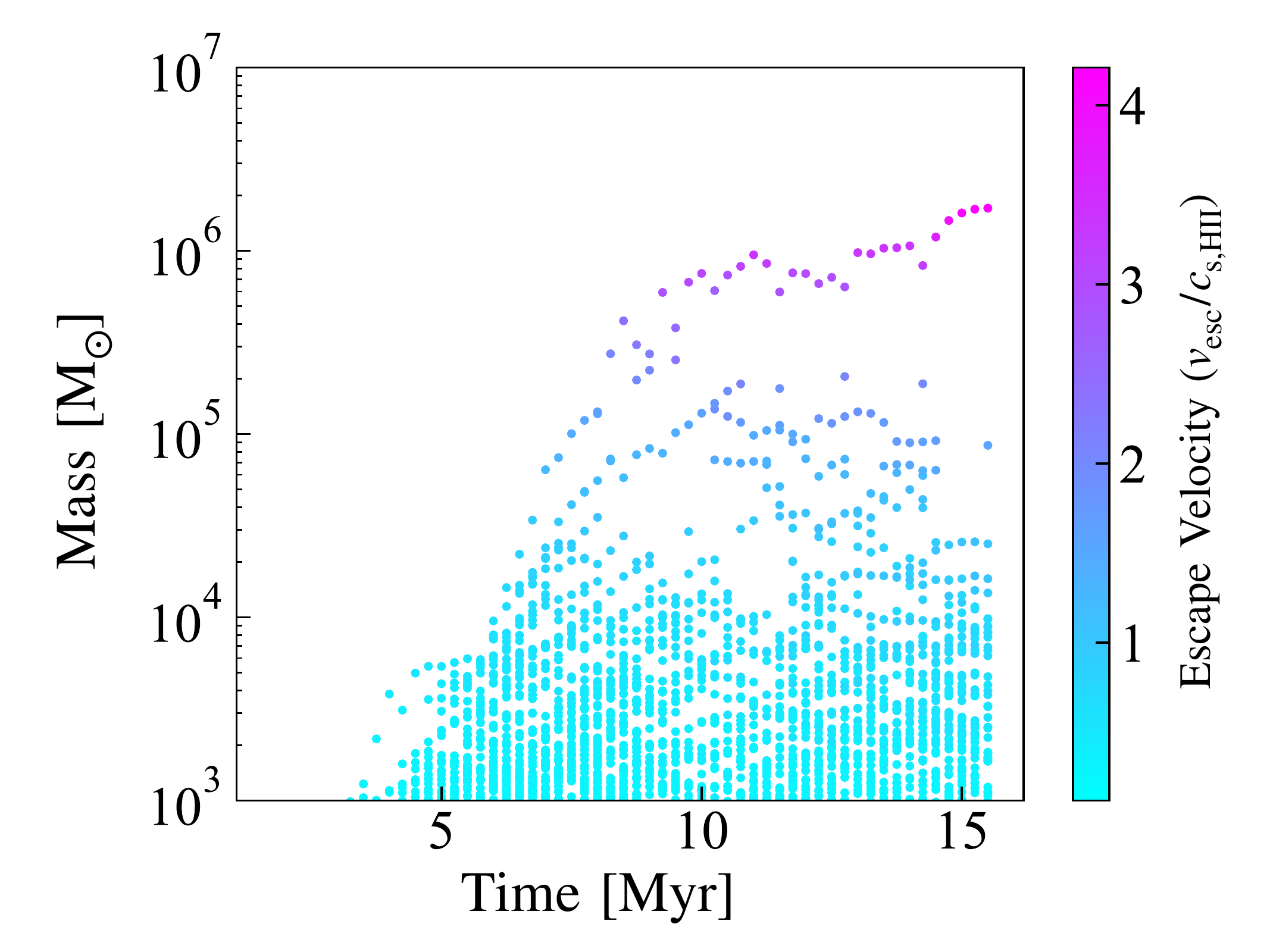}
\caption{Evolution of clump mass and escape velocity for the V100D10 model, the case of fast ($v_0=100\ \mathrm{km\ s^{-1}}$) collisions of dense ($n_0=10\ \mathrm{cm^{-3}}$) gas. The representation follows the same format as in Figure \ref{fig:clmp-vesc}.
\label{fig:v100d10vesc1}}
\end{figure}

We now explore the outcomes of high-speed collisions at $100~\mathrm{km~s^{-1}}$ involving dense gas ($10~\mathrm{cm^{-3}}$) (referred to as V100D10 in Table \ref{tab:models}).

The red, blue, and green dotted lines in Figure \ref{fig:TotalMass} show the evolution of the total gas mass, stellar mass, and massive star mass in this model, respectively. 
Since the shocked sheet is more compressed by stronger ram pressure of the inflow (eq.~\ref{eq:shockjump-n}), resulting $t_\mathrm{ff,sheet}$ is $\sim 1.8~\mathrm{Myr}$, star formation occurs more rapidly than in the fiducial model (V100D1). 
The shocked sheet becomes more massive due to a larger mass flux than the fiducial model, resulting in active star formation with a higher SFE of $\sim 20\%$ (Figure \ref{fig:totSFE}). 
In this case, as shown in Figure \ref{fig:particl} (green line) that a number of massive star clusters are indeed formed through this active star formation.

Figure \ref{fig:v100d10vesc1} shows the clump mass evolution, with its escape velocity. 
We observe the formation of even massive clumps in this case compared to our fiducial case. 
These clumps are approximately an order of magnitude more massive, with the most massive one reaching $\sim1.6\times10^{6}~\mathrm{M_\odot}$ and ranging about $\sim9~\mathrm{pc}$ in size ($\Sigma \sim 2\times10^4~\mathrm{M_\odot}~\mathrm{pc^{-2}}$). 
The resulting clump mass distributions are presented in the middle panels (green) of Figure \ref{fig:hist_sct} at four different epochs: $t = 5.4$, $7.2$, $9.0$, and $14.4~\mathrm{Myr}$, corresponding to approximately 3, 4, 5, and 8 $t_\mathrm{ff,sheet}$, respectively.
This figure also confirms the formation of more massive gas clumps than the fiducial model at $8~t_\mathrm{ff,sheet}$, indicating that a high-density gas collision is important for the formation of YMCs with $\gtrsim10^6~\mathrm{M_\odot}$.

In \cite{tsuge2021formation}, they showed a positive correlation between the ram pressure of the colliding gas and the YMC mass by comparing several YMC forming regions, concluding that ram pressure is important for determining YMC mass. 
Observationally, 
the YMC mass increases almost proportionately to the ram pressure of colliding gas \citep{tsuge2021formation}. 
This is consistent with our result, which shows that a ten times more massive clump is formed when the ram pressure is 10 times higher than the fiducial model. 
If the ram pressure of the gas inflow is greater, the magnetic field at the preshock region should be larger to balance it, which may help the creation of a more massive clump supported by the strong magnetic field. 
The ram pressure of the inflow is determined by its density and velocity, which is about the galaxies' escape velocity in the case of interacting galaxies. 
The formation of a more massive YMC ($\gtrsim10^7~\mathrm{M_\odot}$) can also be expected if dense gas falls into a more massive galaxy with a larger escape velocity.

\subsection{Lower-velocity ($v_0=15\ \mathrm{km\ s^{-1}}$) gas collisions}\label{sec:slow}

As demonstrated in Section~\ref{sec:fast}, fast collisions of HI gas at velocities around $100~\mathrm{km~s^{-1}}$ leads to the formation of sufficiently large ($\gtrsim 10^5~\mathrm{M_{\odot}}$) and compact ($\sim$ pc) gas clumps. Remarkably, these clumps withstand the feedback effects from the formation of HII regions by massive stars and evolve into young massive clusters (YMCs).

In this section, we examine whether YMC can be formed when the shock velocity is not as high as in the case studied in the previous section. For instance, we consider the case of a superbubble shock. 
The typical shock velocity in this case can be estimated by the following expression: 
\begin{align}
  \begin{split}
    v_{\mathrm{sh}}  \simeq 15&~\mathrm{~km} \mathrm{~s}^{-1} \left( \frac{N_\mathrm{SN}}{20}\right)^{1 / 5}
    \left( \frac{E_\mathrm{SN}}{10^{51}~\mathrm{erg}}\right)^{-2 / 5}  \\
    & 
    \left( \frac{n_0}{0.5~\mathrm{cm^{-3}}}\right)^{-1 / 5}
    \left( \frac{t_\mathrm{age}}{10~\mathrm{Myr}}\right)^{-1 / 5},
  \end{split}
  \label{eq:vSN}
\end{align}
where $N_\mathrm{SN}$ is the number of supernovae, $E_\mathrm{SN}$ the energy released by each supernova, $n_0$ the ISM density, and $t_\mathrm{age}$ the age of the superbubble \citep{Weaver1977,Tomisaka1981}.
In this section, we present simulation outcomes for a shock velocity of $v_0=15~\mathrm{km~s^{-1}}$. 

\subsubsection{Fiducial density case (V15D1)}

\begin{figure}[htbp]
\centering
\includegraphics[width=0.4\textwidth]{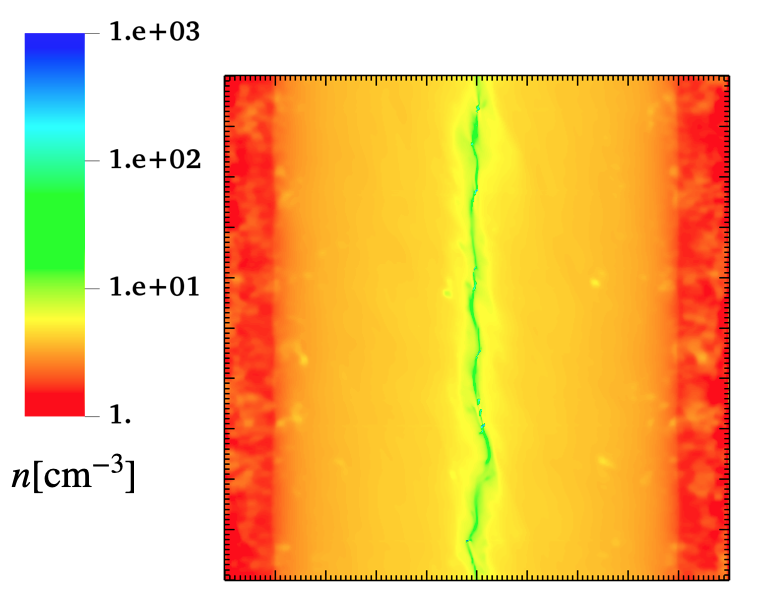}
\caption{Two-dimensional density cross-section of the V15D1 model, the case of slow ($v_0=15\ \mathrm{km\ s^{-1}}$) collisions of fiducial density ($n_0=1\ \mathrm{cm^{-3}}$) gas $t=7.5\  \mathrm{Myr}$ \label{fig:v-slow}. 
}
\end{figure}

\begin{figure}[htbp]
\centering
\includegraphics[width=0.4\textwidth]{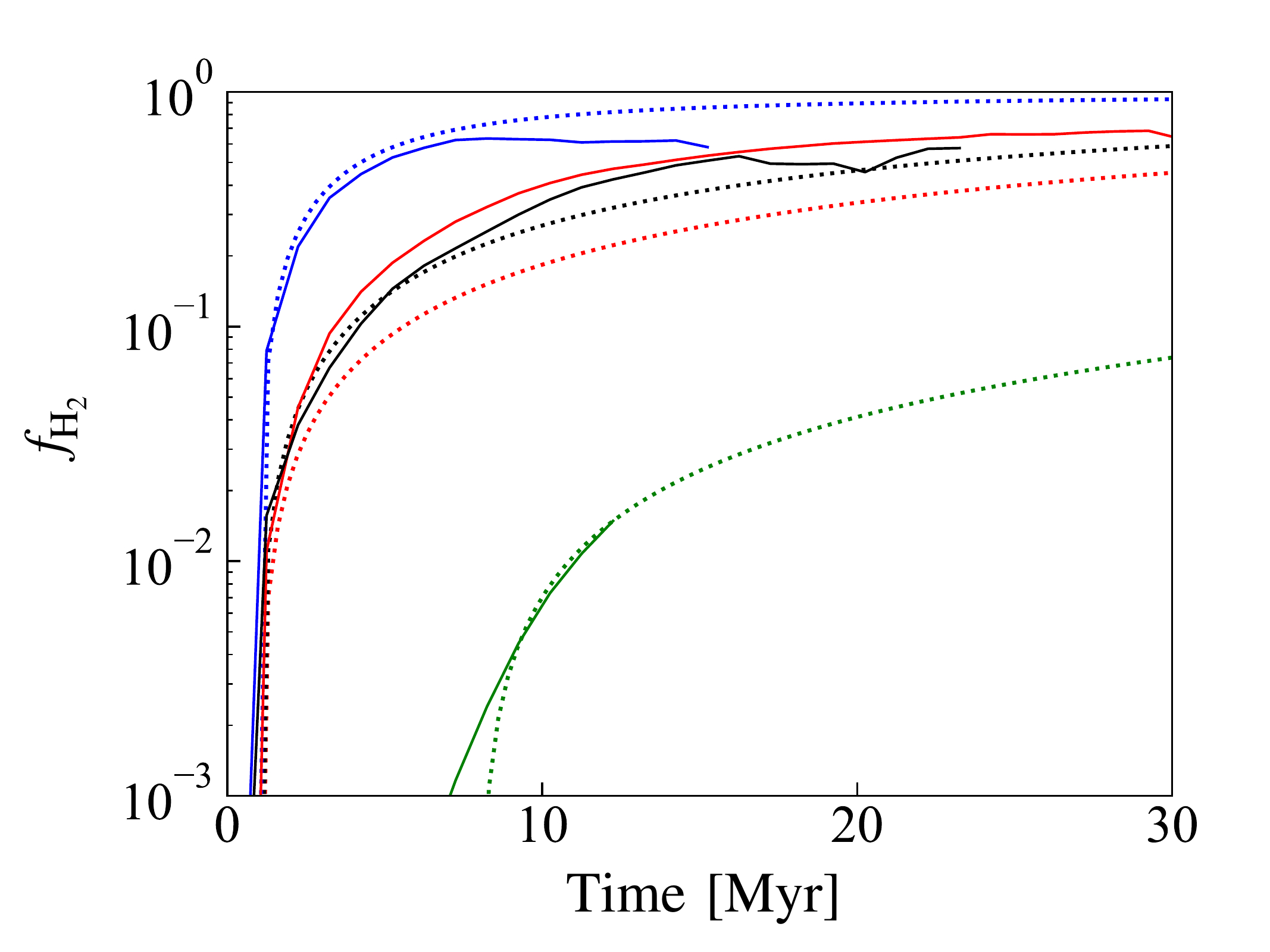}
\caption{Evolution of molecular fraction at shocked gas. The red, blue, green, and black solid lines show the results of models V100D1, V100D10, V15D1, and V15D10, respectively. 
The dotted lines mean the results of eq.~(\ref{eq:h2mol}). 
\label{fig:H2fra} 
}
\end{figure}

We first examine the case where gas collides with the fiducial ISM density $n_0=1\ \mathrm{cm^{-3}}$. 
The density structure at $t=7.5\ \mathrm{Myr}$, depicted in Figure \ref{fig:v-slow}, shows that the slow collision primarily leads to the formation of a low-density shocked layer. 
This agrees with an estimated average number density of the post-shock region, approximately $\bar{n} \sim 5.25\ \mathrm{cm}^{-3}$, calculated using equation (\ref{eq:shockjump-n}). 
In such low-density gas, it takes a prolonged time to form molecules. 
Figure \ref{fig:H2fra} shows $\mathrm{H_2}$ molecular fraction at shocked gas of V100D1 (red), V100D10 (blue), V15D1 (green), and V15D10 (black) models, respectively.  
Note that $f_\mathrm{H_2}$ is defined as the mass fraction of  $\mathrm{H_2}$ among the hydrogen nuclei. 
The result of this model (V15D1, green) indicates a very small molecular fraction ($f_\mathrm{H_2}\lesssim 10^{-2}$) compared to the other models which achieve $f_\mathrm{H_2}\sim 0.5$, indicating that a molecular cloud cannot form in this model. 
This result means star formation cannot occur in this model because molecular clouds, progenitors of star formation, do not form. 

\cite{iwasaki2019early} derived averaged $f_\mathrm{H_2}$ of shocked gas with density inhomogeneity as follows:
\begin{equation}
f_{\mathrm{H}_{2}}(n)=1-\frac{1-e^{-2 k_{\mathrm{H}_{2}} n t_{\mathrm{f}}}}{2 k_{\mathrm{H}_{2}} n t_{\mathrm{f}}}, \label{eq:h2mol}
\end{equation}
where $k_{\mathrm{H}_{2}}=2\times10^{-17}~\mathrm{cm~s^{-1}}$ is the $\mathrm{H_2}$ formation rate assuming that the gas and dust temperatures are $100~\mathrm{K}$ and $10~\mathrm{K}$, respectively \citep{hollenbach1979molecule}.  
This relation is overlaid with dotted lines in Figure \ref{fig:H2fra}. 
Here, we derive the number density of preshock gas $n$ by substituting the model parameters into eq.~(\ref{eq:shockjump-n}). 
Note that the initiation time of molecular formation $t_\mathrm{0}$   ($t_\mathrm{f}=t'_\mathrm{f}-t_\mathrm{0}$) is set to $\sim1~\mathrm{Myr}$ to fit the results of V100D1, V100D10, and V15D10. 
This corresponds to the cooling timescale $\sim1~\mathrm{Myr}$, when hydrogen molecules formation rate becomes sufficiently large owing to temperature drop \citep{hollenbach1979molecule}. 
For this model (V15D1), we apply $t_\mathrm{0}\sim8~\mathrm{Myr}$, at which UV shielding starts to operate. 
From eq.~(\ref{eq:h2mol}), the molecular fractions at a certain time, for example, $t'_\mathrm{f}=15~\mathrm{Myr}$ is estimated as follows: $f_\mathrm{H_2,V100D1}=0.27$, $f_\mathrm{H_2,V100D10}=0.85$, $f_\mathrm{H_2,V15D1}=0.02$, and $f_\mathrm{H_2,V15D10}=0.38$. 
In this model (V15D1), because of the low density, the shocked sheet cannot form a molecular cloud efficiently, thus star formation does not occur.
These results indicate that it is difficult for low-density gas ($\sim 1~\mathrm{cm^{-3}}$) to directly induce star formation by the slow collision. 
Notably, this result is consistent with the molecular cloud formation scenario in a galaxy, in which molecular clouds are formed by multiple shock wave compression of WNM \citep[e.g.,][]{inoue2009two}.

\subsubsection{High density case (V15D10) }

\begin{figure}[htbp]
\centering
\includegraphics[width=0.4\textwidth]{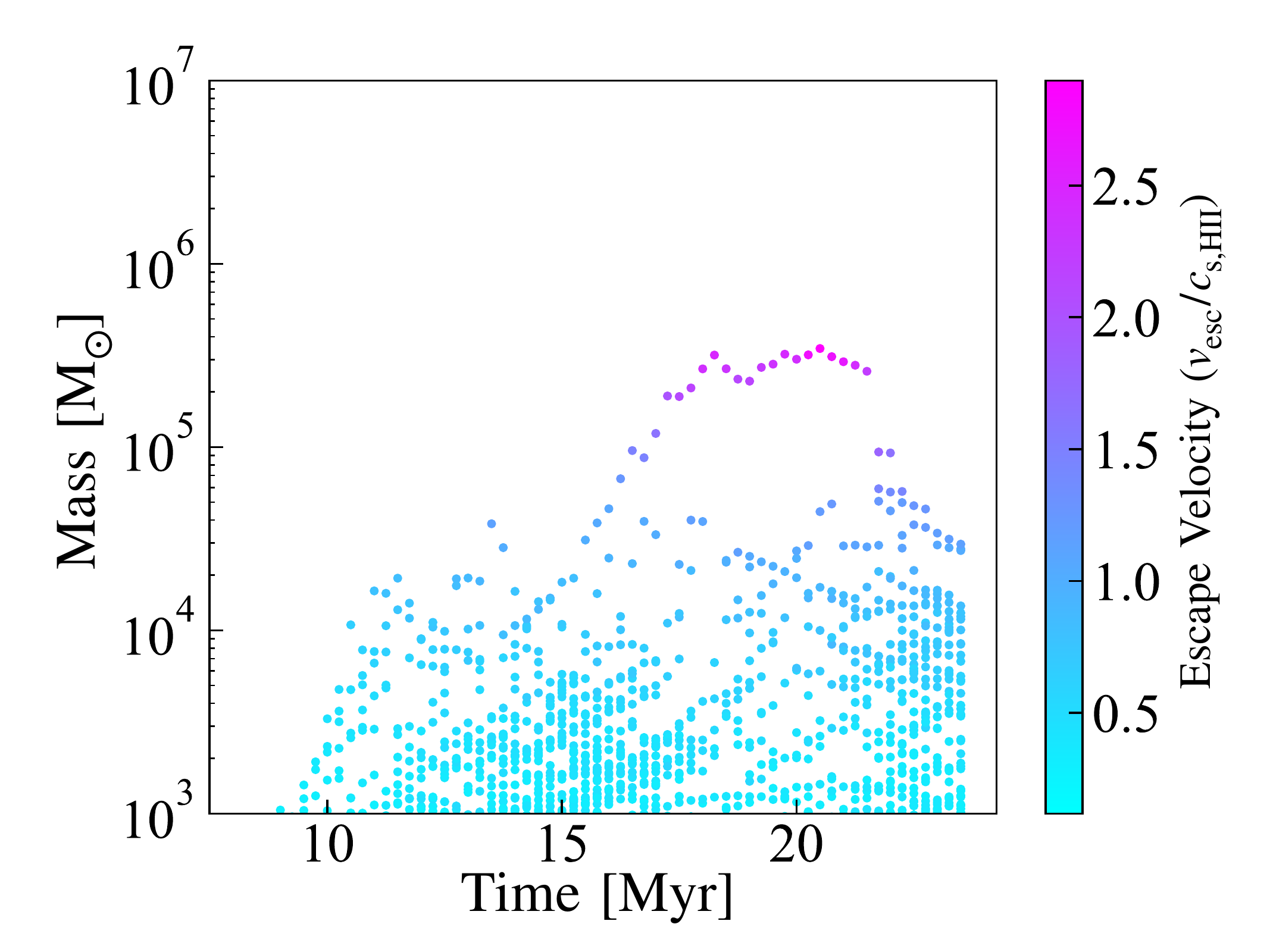}
\caption{Same as Figure \ref{fig:v100d10vesc1} 
but for the V15D10, which is the case of slow ($v_0=15\ \mathrm{km\ s^{-1}}$) collisions of dense ($n_0=10\ \mathrm{cm^{-3}}$) gas. 
\label{fig:v15d10vesc1}}
\end{figure}

Lastly, we examine the case of high-density gas collisions ($10~\mathrm{cm^{-3}}$) with low velocity ($15~\mathrm{km~s^{-1}}$). 
In this case, molecular hydrogen forms on the postshock region, and star formation occurs in contrast to the lower-density case of  V15D1. 
From eq.~(\ref{eq:shockjump-n}), the density of the shocked sheet is about 1.5 times larger than the fiducial model ($\sim55~\mathrm{cm^{-3}}$), resulting in $t_\mathrm{ff,sheet}=4.7~\mathrm{Myr}$. 
Therefore, the time at which star formation occurs is intermediate between the two high-velocity models of V100D1 and V100D10 (see dashed lines in Figure \ref{fig:TotalMass}). 
The final SFE of this model is $\sim10\%$ (the dashed line in Figure \ref{fig:totSFE}) and the mass distribution of formed star clusters is shown by the blue line in Figure \ref{fig:particl}. 

The evolution of clump mass is depicted in Figure \ref{fig:v15d10vesc1}, where the clump escape velocity is also indicated by color. We see that a significant gas clump with mass exceeding $10^5~\mathrm{M_\odot}$ forms around $17~\mathrm{Myr}$. Eventually, this clump reaches a mass $\sim4 \times10^5~\mathrm{M_\odot}$ and a size $L\sim 7~\mathrm{pc}$ by $t=20~\mathrm{Myr}$ ($\Sigma \sim 1\times10^4~\mathrm{M_\odot}~\mathrm{pc^{-2}}$), resulting in escape velocity of $\sim25~\mathrm{km~s^{-1}}$. 
Here again, clumps with lower escape velocities yield disruption from feedback, while those with higher escape velocities persist over an extended period. 
The bottom panels (blue) in Figure \ref{fig:hist_sct} display the mass distribution of the formed clumps at various epochs: $t = 14$, $19$, and $24.5~\mathrm{Myr}$, corresponding to 3, 4, and 5 times the free-fall time. 
This suggests that even at $v_0=15~\mathrm{km~s^{-1}}$, the collision can form a massive shocked sheet (same as the fiducial model), thereby promoting the formation of YMCs.

However, this result does not directly indicate that YMC can be easily formed in galactic super-bubbles. 
Our results suggest that an extended period of gas compression, exceeding $10~\mathrm{Myr}$, is required for the massive YMC progenitor clump formation (see, Figure \ref{fig:v15d10vesc1}). 
For instance, considering a superbubble generated by supernova explosions, the lifetime of the shell having expanding velocity of $\sim 15~\mathrm{km~s^{-1}}$ can be estimated from equation (\ref{eq:vSN}) as follows:
\begin{align}
  \begin{split}
    & t \simeq 0.5~\mathrm{Myr}  \\
    & \left( \frac{v_{\mathrm{sh}} }{15~\mathrm{~km} \mathrm{~s}^{-1}} \right)^{-5}
      \left( \frac{N_\mathrm{SN}}{20}\right)
\left( \frac{E_\mathrm{SN}}{10^{51}~\mathrm{erg}}\right)^{- 2}
    \left( \frac{n_0}{10~\mathrm{cm^{-3}}}\right)^{-1}.
  \end{split}
  \label{eq:tSN}
\end{align}
To allow sufficient compression for YMC formation ($\sim 10~\mathrm{Myr}$), a superbubble must involve an unrealistically huge number of explosions $N_\mathrm{SN}\gtrsim400$. This implies that the formation of YMCs via compression by a typical superbubble in a galaxy presents a significant challenge. 
In addition, we assume here that the $10~\mathrm{cm^{-3}}$ gas is spread out about $150~\mathrm{pc}$, but the size of the interstellar cloud in the galaxy is typically not so large. This assumption also makes YMC formation by superbubbles more difficult.

\section{Summary \& Discussion}
We performed numerical simulations to explore the formation of star clusters via HI gas collision while accounting for the photoionization effects due to the emergence of massive stars. For this purpose, we considered photoionization feedback in an approximate fashion by elevating the gas temperature within the Strömgren radius to $10^4\ \mathrm{K}$. 
Our findings can be summarized as follows:
\begin{itemize}
    \item[1.] Even in the presence of photoionization feedback, massive ($> 10^5\ \mathrm{M_\odot}$) and compact ($\sim \mathrm{pc}$) gas clumps can be produced through HI gas ($1~\mathrm{cm^{-3}}$) collisions with fast velocity $v=100~\mathrm{km~s^{-1}}$. The resulting clumps are sufficiently compact, with escape velocities surpassing the sound speed of the HII region, indicating tight gravitational binding. Consequently, these circumstances promote efficient star formation within the massive gas clumps, ultimately leading to their evolution into young massive clusters (YMCs). Conversely, when massive stars emerge within low-mass clumps, feedback effects induce clump evaporation. 
        \item[2.] 
        When considering dense gas inflows ($10~\mathrm{cm^{-3}}$), fast collisions ($100~\mathrm{km~s^{-1}}$) allow the formation of exceptionally massive gas clumps exceeding $10^6~\mathrm{M_\odot}$.
        This is consistent with the observational fact that regions experiencing higher ram pressure collisions are conducive to the formation of more massive star clusters.  

    \item[3.] In cases of low-velocity ($15~\mathrm{km~s^{-1}}$) gas collisions, such as those induced by supernova shocks, the formation of molecular gas within the shocked layer is inefficient, especially at colliding gas density as low as $1~\mathrm{cm^{-3}}$. 
    This is because the sheet formed by weak collisions has a low density and takes a long time to form molecular gas, resulting in failed star formation. 
    Even in low-velocity collisions, if denser gas ($10~\mathrm{cm^{-3}}$) collide, massive gas clumps can form as long as the compression continues $\gtrsim 10~\mathrm{Myr}$. 
    However, achieving such prolonged compression by supernova-induced shocks within a galaxy hardly happens, indicating the preference for large-scale gas collisions arising from galactic interactions.
   
\end{itemize}

In this paper, we have considered only photoionization as feedback from massive stars, as it is the primary feedback mechanism in YMC formation \citep{fukushima2022far}. 
Another major form of stellar feedback, supernova explosions, is not expected to have a significant impact on preventing the evolution of our massive clumps into YMCs for the following reasons.

The effects of supernova explosions begin to take place after the lifetime of massive stars, which depends on their mass; it is $8~\mathrm{Myr}$ for a $20~\mathrm{M_\odot}$ star and $3~\mathrm{Myr}$ for a $120~\mathrm{M_\odot}$ star \citep{pagel2009nucleosynthesis}. Since stars with masses of $\gtrsim 100\ \mathrm{M_\odot}$ are very rare considering the IMF, we adopt $8.2~\mathrm{Myr}$ as the typical timescale for supernova explosions in our discussion.
This timescale should be compared with the duration of star formation in massive gas clumps, which is approximately $2-3$ times the free-fall time of the clump \citep{fukushima2024impacts}, amounting to around $0.4-0.6~\mathrm{Myr}$ in our fiducial model. This duration is considerably shorter than the onset time of supernova feedback.
This suggests that once massive gas clumps are formed, they can potentially form massive star clusters before supernova feedback becomes operative.

Another possibility is that the growth of small gas clumps into massive clumps takes a longer time, and if massive stars form within these small clumps before they grow into massive clumps, it could impact the formation of massive clumps and the YMCs within them. In our simulation, the time required for the formation of a massive clump of $\sim 10^5~\mathrm{M_\odot}$ is around $3-5$ times the free-fall time of their progenitor shocked sheet-like clouds ($t_\mathrm{ff,sheet}$) (see Figure \ref{fig:hist_sct}). In models with high sheet densities (i.e., cases with large shock velocities and initial densities, such as V100D10), this corresponds to $3.6~\mathrm{Myr}$, which is shorter than the typical lifetime of massive stars ($8.2~\mathrm{Myr}$).
On the other hand, in models with lower sheet densities such as V100D1 and V15D10, this timescale becomes as long as $12.0~\mathrm{Myr}$ and $9.4~\mathrm{Myr}$, respectively, somewhat exceeding the characteristic massive star lifetime. Therefore, if massive stars form at the very beginning in low-mass clumps before the formation of massive clumps, supernova explosions could prevent the formation of massive clumps and the YMCs within them.
However, observationally, it is known that massive stars are more likely to form in larger clumps rather than in small clumps, and they tend to form during the later stages of star formation within these large clumps \citep{kumar2020unifying}. In such cases, the influence of SN feedback on the formation of massive gas clumps would be limited. We plan to investigate the formation of massive star clusters while accounting for the effects of supernova feedback in future work.

\begin{acknowledgments}
We are grateful to Shu-ichiro Inutsuka for his helpful comments and suggestions.
Numerical computations were conducted on Cray XC50 {\tt Aterui II} in Oshu City at the Center for Computational Astrophysics (CfCA) of the National Astronomical Observatory of Japan. 
The computation was also carried out using the JHPCN Joint Research Projects on supercomputer {\tt Flow} at the Information Technology Center, Nagoya University. 
This work is supported by grant-in-aid from the Ministry of Education, Culture, Sports, Science, and Technology (MEXT) of Japan, Grant No. 23H00129 (TI) and 22H00149 (KO).  
\end{acknowledgments}

\bibliography{sample631}{}

\begin{thebibliography}{}
\expandafter\ifx\csname natexlab\endcsname\relax\def\natexlab#1{#1}\fi
\providecommand{\url}[1]{\href{#1}{#1}}
\providecommand{\dodoi}[1]{doi:~\href{http://doi.org/#1}{\nolinkurl{#1}}}
\providecommand{\doeprint}[1]{\href{http://ascl.net/#1}{\nolinkurl{http://ascl.net/#1}}}
\providecommand{\doarXiv}[1]{\href{https://arxiv.org/abs/#1}{\nolinkurl{https://arxiv.org/abs/#1}}}

\bibitem[{{Adamo} {et~al.}(2024){Adamo}, {Bradley}, {Vanzella}, {Claeyssens}, {Welch}, {Diego}, {Mahler}, {Oguri}, {Sharon}, {Abdurro'uf}, {Hsiao}, {Messa}, {Zackrisson}, {Brammer}, {Coe}, {Kokorev}, {Ricotti}, {Zitrin}, {Fujimoto}, {Inoue}, {Resseguier}, {Rigby}, {Jim{\'e}nez-Teja}, {Windhorst}, \& {Xu}}]{2024arXiv240103224A}
{Adamo}, A., {Bradley}, L.~D., {Vanzella}, E., {et~al.} 2024, arXiv e-prints, arXiv:2401.03224, \dodoi{10.48550/arXiv.2401.03224}

\bibitem[{Ballesteros-Paredes {et~al.}(1999)Ballesteros-Paredes, Hartmann, \& V{\'a}zquez-Semadeni}]{ballesteros1999turbulent}
Ballesteros-Paredes, J., Hartmann, L., \& V{\'a}zquez-Semadeni, E. 1999, \apj, 527, 285

\bibitem[{Banerjee {et~al.}(2009)Banerjee, V{\'a}zquez-Semadeni, Hennebelle, \& Klessen}]{banerjee2009clump}
Banerjee, R., V{\'a}zquez-Semadeni, E., Hennebelle, P., \& Klessen, R. 2009, \mnras, 398, 1082

\bibitem[{Bekki \& Chiba(2007)}]{Bekki2007}
Bekki, K., \& Chiba, M. 2007, \apj, 665, 1164

\bibitem[{Clarke(1996)}]{clarke1996consistent}
Clarke, D.~A. 1996, \apj, 457, 291

\bibitem[{Col{\'\i}n {et~al.}(2013)Col{\'\i}n, V{\'a}zquez-Semadeni, \& G{\'o}mez}]{colin2013molecular}
Col{\'\i}n, P., V{\'a}zquez-Semadeni, E., \& G{\'o}mez, G.~C. 2013, \mnras, 435, 1701

\bibitem[{Dale {et~al.}(2012)Dale, Ercolano, \& Bonnell}]{dale2012ionizing}
Dale, J., Ercolano, B., \& Bonnell, I. 2012, Monthly Notices of the Royal Astronomical Society, 424, 377

\bibitem[{Dobbs {et~al.}(2020)Dobbs, Liow, \& Rieder}]{dobbs2020formation}
Dobbs, C., Liow, K., \& Rieder, S. 2020, Monthly Notices of the Royal Astronomical Society: Letters, 496, L1

\bibitem[{Dobbs \& Wurster(2021)}]{dobbs2021properties}
Dobbs, C., \& Wurster, J. 2021, Monthly Notices of the Royal Astronomical Society, 502, 2285

\bibitem[{Elmegreen \& Efremov(1997)}]{elmegreen1997universal}
Elmegreen, B.~G., \& Efremov, Y.~N. 1997, \apj, 480, 235

\bibitem[{Fujii \& Portegies~Zwart(2016)}]{fujii2016formation}
Fujii, M., \& Portegies~Zwart, S. 2016, \apj, 817, 4

\bibitem[{Fujii {et~al.}(2022)Fujii, Hattori, Wang, Hirai, Kumamoto, Shimajiri, \& Saitoh}]{fujii2022sirius}
Fujii, M.~S., Hattori, K., Wang, L., {et~al.} 2022, Monthly Notices of the Royal Astronomical Society, 514, 43

\bibitem[{Fujii \& Portegies~Zwart(2015)}]{Fujii2015}
Fujii, M.~S., \& Portegies~Zwart, S.~P. 2015, Proceedings of the International Astronomical Union, 12, 25

\bibitem[{Fujimoto \& Noguchi(1990)}]{fujimoto1990asymmetric}
Fujimoto, M., \& Noguchi, M. 1990, \pasj, 42, 505

\bibitem[{Fukui {et~al.}(2017)Fukui, Tsuge, Sano, Bekki, Yozin, Tachihara, \& Inoue}]{fukui2017formation}
Fukui, Y., Tsuge, K., Sano, H., {et~al.} 2017, \pasj, 69, L5

\bibitem[{Fukushima \& Yajima(2022)}]{fukushima2022far}
Fukushima, H., \& Yajima, H. 2022, Monthly Notices of the Royal Astronomical Society, 511, 3346

\bibitem[{Fukushima \& Yajima(2024)}]{fukushima2024impacts}
---. 2024, arXiv preprint arXiv:2404.10535

\bibitem[{Fukushima {et~al.}(2020)Fukushima, Yajima, Sugimura, Hosokawa, Omukai, \& Matsumoto}]{fukushima2020star}
Fukushima, H., Yajima, H., Sugimura, K., {et~al.} 2020, Monthly Notices of the Royal Astronomical Society, 497, 3830

\bibitem[{Furuta {et~al.}(2021)Furuta, Kaneda, Kokusho, Nakajima, Fukui, \& Tsuge}]{furuta2021three}
Furuta, T., Kaneda, H., Kokusho, T., {et~al.} 2021, Publications of the Astronomical Society of Japan, 73, 864

\bibitem[{Furuta {et~al.}(2022)Furuta, Kaneda, Kokusho, Nakajima, Fukui, \& Tsuge}]{furuta2022three}
---. 2022, Publications of the Astronomical Society of Japan, 74, 639

\bibitem[{Gaensler {et~al.}(2005)Gaensler, Haverkorn, Staveley-Smith, Dickey, McClure-Griffiths, Dickel, \& Wolleben}]{gaensler2005magnetic}
Gaensler, B.~M., Haverkorn, M., Staveley-Smith, L., {et~al.} 2005, Sci, 307, 1610

\bibitem[{{Garcia} {et~al.}(2023){Garcia}, {Ricotti}, {Sugimura}, \& {Park}}]{2023MNRAS.522.2495G}
{Garcia}, F. A.~B., {Ricotti}, M., {Sugimura}, K., \& {Park}, J. 2023, \mnras, 522, 2495, \dodoi{10.1093/mnras/stad1092}

\bibitem[{Gong \& Ostriker(2012)}]{gong2012implementation}
Gong, H., \& Ostriker, E.~C. 2012, The Astrophysical Journal Supplement Series, 204, 8

\bibitem[{Gonz{\'a}lez-Samaniego \& Vazquez-Semadeni(2020)}]{gonzalez2020effect}
Gonz{\'a}lez-Samaniego, A., \& Vazquez-Semadeni, E. 2020, arXiv preprint arXiv:2003.12711

\bibitem[{Grudi{\'c} {et~al.}(2018)Grudi{\'c}, Hopkins, Faucher-Giguere, Quataert, Murray, \& Kere{\v{s}}}]{grudic2018feedback}
Grudi{\'c}, M.~Y., Hopkins, P.~F., Faucher-Giguere, C.-A., {et~al.} 2018, Monthly Notices of the Royal Astronomical Society, 475, 3511

\bibitem[{Guszejnov {et~al.}(2022)Guszejnov, Markey, Offner, Grudi{\'c}, Faucher-Gigu{\`e}re, Rosen, \& Hopkins}]{guszejnov2022cluster}
Guszejnov, D., Markey, C., Offner, S.~S., {et~al.} 2022, Monthly Notices of the Royal Astronomical Society, 515, 167

\bibitem[{Habing(1968)}]{habing1968interstellar}
Habing, H. 1968, BAN, 19, 421

\bibitem[{Hartmann {et~al.}(2001)Hartmann, Ballesteros-Paredes, \& Bergin}]{hartmann2001rapid}
Hartmann, L., Ballesteros-Paredes, J., \& Bergin, E.~A. 2001, \apj, 562, 852

\bibitem[{He {et~al.}(2019)He, Ricotti, \& Geen}]{he2019simulating}
He, C.-C., Ricotti, M., \& Geen, S. 2019, Monthly Notices of the Royal Astronomical Society, 489, 1880

\bibitem[{He {et~al.}(2020)He, Ricotti, \& Geen}]{he2020simulating}
---. 2020, Monthly Notices of the Royal Astronomical Society, 492, 4858

\bibitem[{Heitsch {et~al.}(2005)Heitsch, Burkert, Hartmann, Slyz, \& Devriendt}]{heitsch2005formation}
Heitsch, F., Burkert, A., Hartmann, L.~W., Slyz, A.~D., \& Devriendt, J.~E. 2005, \apjl, 633, L113

\bibitem[{Heitsch {et~al.}(2008)Heitsch, Hartmann, Slyz, Devriendt, \& Burkert}]{heitsch2008cooling}
Heitsch, F., Hartmann, L.~W., Slyz, A.~D., Devriendt, J.~E., \& Burkert, A. 2008, \apj, 674, 316

\bibitem[{Heitsch {et~al.}(2006)Heitsch, Slyz, Devriendt, Hartmann, \& Burkert}]{heitsch2006birth}
Heitsch, F., Slyz, A.~D., Devriendt, J.~E., Hartmann, L.~W., \& Burkert, A. 2006, \apj, 648, 1052

\bibitem[{Heitsch {et~al.}(2009)Heitsch, Stone, \& Hartmann}]{heitsch2009effects}
Heitsch, F., Stone, J.~M., \& Hartmann, L.~W. 2009, \apj, 695, 248

\bibitem[{Hennebelle {et~al.}(2008)Hennebelle, Banerjee, V{\'a}zquez-Semadeni, Klessen, \& Audit}]{hennebelle2008warm}
Hennebelle, P., Banerjee, R., V{\'a}zquez-Semadeni, E., Klessen, R., \& Audit, E. 2008, \aap, 486, L43

\bibitem[{Hirai {et~al.}(2021)Hirai, Fujii, \& Saitoh}]{hirai2021sirius}
Hirai, Y., Fujii, M.~S., \& Saitoh, T.~R. 2021, Publications of the Astronomical Society of Japan, 73, 1036

\bibitem[{Hollenbach \& McKee(1979)}]{hollenbach1979molecule}
Hollenbach, D., \& McKee, C.~F. 1979, \apjs, 41, 555

\bibitem[{Inoue \& Inutsuka(2008)}]{inoue2008two}
Inoue, T., \& Inutsuka, S. 2008, \apj, 687, 303

\bibitem[{Inoue \& Inutsuka(2009)}]{inoue2009two}
---. 2009, \apj, 704, 161

\bibitem[{Inoue \& Inutsuka(2012)}]{inoue2012formation}
---. 2012, \apj, 759, 35

\bibitem[{Inoue \& Inutsuka(2016)}]{Inoue2016}
---. 2016, \apj, 833, 10

\bibitem[{Inoue \& Omukai(2015)}]{Inoue2015}
Inoue, T., \& Omukai, K. 2015, \aj, 805

\bibitem[{Iwasaki \& Tomida(2022)}]{iwasaki2022universal}
Iwasaki, K., \& Tomida, K. 2022, The Astrophysical Journal, 934, 174

\bibitem[{Iwasaki {et~al.}(2019)Iwasaki, Tomida, Inoue, \& Inutsuka}]{iwasaki2019early}
Iwasaki, K., Tomida, K., Inoue, T., \& Inutsuka, S.-i. 2019, The Astrophysical Journal, 873, 6

\bibitem[{Kim {et~al.}(2018)Kim, Kim, \& Ostriker}]{kim2018modeling}
Kim, J.-G., Kim, W.-T., \& Ostriker, E.~C. 2018, The Astrophysical Journal, 859, 68

\bibitem[{Kim {et~al.}(2021)Kim, Ostriker, \& Filippova}]{kim2021star}
Kim, J.-G., Ostriker, E.~C., \& Filippova, N. 2021, The Astrophysical Journal, 911, 128

\bibitem[{Kim {et~al.}(1998)Kim, Staveley-Smith, Dopita, Freeman, Sault, Kesteven, \& McConnell}]{kim1998hi}
Kim, S., Staveley-Smith, L., Dopita, M.~A., {et~al.} 1998, \apj, 503, 674

\bibitem[{Kim {et~al.}(2003)Kim, Staveley-Smith, Dopita, Sault, Freeman, Lee, \& Chu}]{kim2003neutral}
---. 2003, \apjs, 148, 473

\bibitem[{Kobayashi {et~al.}(2020)Kobayashi, Inoue, Inutsuka, Tomida, Iwasaki, \& Tanaka}]{kobayashi2020bimodal}
Kobayashi, M.~I., Inoue, T., Inutsuka, S.-i., {et~al.} 2020, The Astrophysical Journal, 905, 95

\bibitem[{Kobayashi {et~al.}(2022)Kobayashi, Inoue, Tomida, Iwasaki, \& Nakatsugawa}]{kobayashi2022nature}
Kobayashi, M.~I., Inoue, T., Tomida, K., Iwasaki, K., \& Nakatsugawa, H. 2022, The Astrophysical Journal, 930, 76

\bibitem[{Kobayashi {et~al.}(2023)Kobayashi, Iwasaki, Tomida, Inoue, Omukai, \& Tokuda}]{kobayashi2023metallicity}
Kobayashi, M.~I., Iwasaki, K., Tomida, K., {et~al.} 2023, The Astrophysical Journal, 954, 38

\bibitem[{Koyama \& Inutsuka(2000)}]{Koyama2000}
Koyama, H., \& Inutsuka, S. 2000, \apj, 532, 980

\bibitem[{Koyama \& Inutsuka(2001)}]{Koyama2001}
---. 2001, \apj, 564, L97

\bibitem[{Kroupa(2001)}]{kroupa2001variation}
Kroupa, P. 2001, Monthly Notices of the Royal Astronomical Society, 322, 231

\bibitem[{Krumholz {et~al.}(2011)Krumholz, Dekel, \& McKee}]{krumholz2011universal}
Krumholz, M.~R., Dekel, A., \& McKee, C.~F. 2011, \apj, 745, 69

\bibitem[{Kumar {et~al.}(2020)Kumar, Palmeirim, Arzoumanian, \& Inutsuka}]{kumar2020unifying}
Kumar, M., Palmeirim, P., Arzoumanian, D., \& Inutsuka, S. 2020, Astronomy \& Astrophysics, 642, A87

\bibitem[{Lanz \& Hubeny(2003)}]{lanz2003grid}
Lanz, T., \& Hubeny, I. 2003, The Astrophysical Journal Supplement Series, 146, 417

\bibitem[{Longmore {et~al.}(2014)Longmore, Kruijssen, Bastian, Bally, Rathborne, Testi, Stolte, Dale, Bressert, \& Alves}]{longmore2014formation}
Longmore, S.~N., Kruijssen, J.~D., Bastian, N., {et~al.} 2014, Protostars and Planets VI, 1, 291

\bibitem[{Maeda {et~al.}(2021)Maeda, Inoue, \& Fukui}]{maeda2021formation}
Maeda, R., Inoue, T., \& Fukui, Y. 2021, The Astrophysical Journal, 908, 2

\bibitem[{Maeda {et~al.}(2024)Maeda, Inoue, \& Inutsuka}]{maeda2024improved}
Maeda, R., Inoue, T., \& Inutsuka, S.-i. 2024, Monthly Notices of the Royal Astronomical Society, 527, 471

\bibitem[{Miyama {et~al.}(1987)Miyama, Narita, \& Hayashi}]{miyama1987fragmentation}
Miyama, S.~M., Narita, S., \& Hayashi, C. 1987, PThPh, 78, 1273

\bibitem[{{Osterbrock} \& {Ferland}(2006)}]{2006agna.book.....O}
{Osterbrock}, D.~E., \& {Ferland}, G.~J. 2006, {Astrophysics of gaseous nebulae and active galactic nuclei}

\bibitem[{Pagel(2009)}]{pagel2009nucleosynthesis}
Pagel, B. E.~J. 2009, Nucleosynthesis and chemical evolution of galaxies (Cambridge University Press)

\bibitem[{Portegies~Zwart {et~al.}(2010)Portegies~Zwart, McMillan, \& Gieles}]{portegies2010young}
Portegies~Zwart, S.~F., McMillan, S.~L., \& Gieles, M. 2010, ARA\&A, 48, 431

\bibitem[{Press {et~al.}(1986)Press, Teukolsky, Vetterling, \& Flannery}]{press1986numerical}
Press, W.~H., Teukolsky, S.~A., Vetterling, W.~T., \& Flannery, B.~P. 1986, Numerical recipes in Fortran 77,  Cambridge university press New York

\bibitem[{Sano {et~al.}(1999)Sano, Inutsuka, \& Miyama}]{sano1999higher}
Sano, T., Inutsuka, S., \& Miyama, S. 1999, in Numerical Astrophysics (Springer), 383--386

\bibitem[{Staveley-Smith(1997)}]{staveley1997hi}
Staveley-Smith, L. 1997, PASA, 14, 111

\bibitem[{{Sugimura} {et~al.}(2024){Sugimura}, {Ricotti}, {Park}, {Garcia}, \& {Yajima}}]{2024arXiv240304824S}
{Sugimura}, K., {Ricotti}, M., {Park}, J., {Garcia}, F. A.~B., \& {Yajima}, H. 2024, arXiv e-prints, arXiv:2403.04824, \dodoi{10.48550/arXiv.2403.04824}

\bibitem[{{Tachihara} {et~al.}(2018){Tachihara}, {Gratier}, {Sano}, {Tsuge}, {Miura}, {Muraoka}, \& {Fukui}}]{tachihara2018}
{Tachihara}, K., {Gratier}, P., {Sano}, H., {et~al.} 2018, \pasj, 70, S52, \dodoi{10.1093/pasj/psy020}

\bibitem[{{Tomisaka} {et~al.}(1981){Tomisaka}, {Habe}, \& {Ikeuchi}}]{Tomisaka1981}
{Tomisaka}, K., {Habe}, A., \& {Ikeuchi}, S. 1981, \apss, 78, 273, \dodoi{10.1007/BF00648941}

\bibitem[{Tsuge {et~al.}(2021{\natexlab{a}})Tsuge, Fukui, Tachihara, Sano, Tokuda, Ueda, Iono, \& Finn}]{tsuge2021formation}
Tsuge, K., Fukui, Y., Tachihara, K., {et~al.} 2021{\natexlab{a}}, Publications of the Astronomical Society of Japan, 73, S35

\bibitem[{Tsuge {et~al.}(2021{\natexlab{b}})Tsuge, Tachihara, Fukui, Sano, Tokuda, Ueda, \& Iono}]{tsuge2021formation2}
Tsuge, K., Tachihara, K., Fukui, Y., {et~al.} 2021{\natexlab{b}}, Publications of the Astronomical Society of Japan, 73, 417

\bibitem[{Tsuge {et~al.}(2019)Tsuge, Sano, Tachihara, Yozin, Bekki, Inoue, Mizuno, Kawamura, Onishi, \& Fukui}]{tsuge2019formation}
Tsuge, K., Sano, H., Tachihara, K., {et~al.} 2019, \apj, 871, 44

\bibitem[{Tsuge {et~al.}(2024)Tsuge, Sano, Tachihara, Bekki, Tokuda, Inoue, Mizuno, Kawamura, Onishi, \& Fukui}]{tsuge2024high}
---. 2024, arXiv preprint arXiv:2405.05046

\bibitem[{Van~Leer(1997)}]{van1997flux}
Van~Leer, B. 1997, in Upwind and High-Resolution Schemes (Springer), 80--89

\bibitem[{Vanzella {et~al.}(2022)Vanzella, Castellano, Bergamini, Treu, Mercurio, Scarlata, Rosati, Grillo, Acebron, Caminha, {et~al.}}]{vanzella2022early}
Vanzella, E., Castellano, M., Bergamini, P., {et~al.} 2022, The Astrophysical Journal Letters, 940, L53

\bibitem[{Vanzella {et~al.}(2023{\natexlab{a}})Vanzella, Claeyssens, Welch, Adamo, Coe, Diego, Mahler, Khullar, Kokorev, Oguri, {et~al.}}]{vanzella2023jwst}
Vanzella, E., Claeyssens, A., Welch, B., {et~al.} 2023{\natexlab{a}}, The Astrophysical Journal, 945, 53

\bibitem[{Vanzella {et~al.}(2023{\natexlab{b}})Vanzella, Loiacono, Bergamini, Me{\v{s}}tri{\'c}, Castellano, Rosati, Meneghetti, Grillo, Calura, Mignoli, {et~al.}}]{vanzella2023extremely}
Vanzella, E., Loiacono, F., Bergamini, P., {et~al.} 2023{\natexlab{b}}, Astronomy \& Astrophysics, 678, A173

\bibitem[{V{\'a}zquez-Semadeni {et~al.}(2007)V{\'a}zquez-Semadeni, G{\'o}mez, Jappsen, Ballesteros-Paredes, Gonz{\'a}lez, \& Klessen}]{vazquez2007molecular}
V{\'a}zquez-Semadeni, E., G{\'o}mez, G.~C., Jappsen, A.~K., {et~al.} 2007, \apj, 657, 870

\bibitem[{V{\'a}zquez-Semadeni {et~al.}(2017)V{\'a}zquez-Semadeni, Gonz{\'a}lez-Samaniego, \& Col{\'\i}n}]{vazquez2017hierarchical}
V{\'a}zquez-Semadeni, E., Gonz{\'a}lez-Samaniego, A., \& Col{\'\i}n, P. 2017, \mnras, 467, 1313

\bibitem[{V{\'a}zquez-Semadeni {et~al.}(2006)V{\'a}zquez-Semadeni, Ryu, Passot, Gonz{\'a}lez, \& Gazol}]{vazquez2006molecular}
V{\'a}zquez-Semadeni, E., Ryu, D., Passot, T., Gonz{\'a}lez, R.~F., \& Gazol, A. 2006, \apj, 643, 245

\bibitem[{{Weaver} {et~al.}(1977){Weaver}, {McCray}, {Castor}, {Shapiro}, \& {Moore}}]{Weaver1977}
{Weaver}, R., {McCray}, R., {Castor}, J., {Shapiro}, P., \& {Moore}, R. 1977, \apj, 218, 377, \dodoi{10.1086/155692}

\end{thebibliography}
\bibliographystyle{aasjournal}

\end{document}